\documentclass[aps,pra,reprint,superscriptaddress,showpacs,longbibliography,floatfix,footnoteinbib]{revtex4-1}
\usepackage{mathtools,amssymb,graphicx,units}  
\usepackage[usenames,dvipsnames]{color}
\usepackage[plainpages=false,pdfpagelabels,colorlinks=true,linkcolor=red,urlcolor=blue,citecolor=blue,pdftitle={Title},pdfauthor={},pdfdisplaydoctitle=true,pdfduplex=DuplexFlipLongEdge]{hyperref}
  
\usepackage{epsf}   
\usepackage{epsfig}
\usepackage{graphics}
\usepackage{amssymb}
\usepackage{hyperref}
\usepackage{color}
\usepackage{subfigure} 
\usepackage{grffile}
\usepackage{bm}
\usepackage{mhchem}
\usepackage{bibentry}
\usepackage{soul}
\begin{document}
 
\title{Magnetic order and transport in a spin-fermion model on a superlattice }

\author{Sabyasachi Tarat}
\affiliation{
 Beijing Computational Science Research Center, Beijing 100193, China
}

\author{Jian Li}
\affiliation{
 Beijing Computational Science Research Center, Beijing 100193, China
}

\author{Richard T. Scalettar}
\affiliation{
Department of Physics, University of California, Davis, California 95616, USA}
\author{Rubem Mondaini}
\affiliation{
 Beijing Computational Science Research Center, Beijing 100193, China
}

% \date{\today}% It is always \today, today,

\begin{abstract}
We consider a spin-fermion model consisting of free electrons coupled to
classical spins, where the latter are embedded in a quasi
one-dimensional superlattice structure consisting of spin blocks
separated by spinless buffers. Using a spiral ansatz for the spins,
we study the effect of the electron mediated
Ruderman-Kittel-Kasuya-Yosida (RKKY) interaction on the $T=0$ ground
state of the system. We find that the RKKY interaction can lead to
ferromagnetic, antiferromagnetic, or intermediate spiral phases for
different system parameters. When the width is much larger than the
length of the individual blocks, the spiral phases are suppressed, and
the ground state oscillates between ferromagnetic and antiferromagnetic
order as the size of the buffer regions is varied. This is accompanied
by a corresponding oscillation in the Drude weight reflecting an increased
conductivity in the ferromagnetic state compared to the
antiferromagnetic one.  These results are reminiscent of classic giant
magnetoresistance phenomena observed in a similar geometry of thin,
sandwiched magnetic and non-magnetic layers. Our analysis provides a
robust framework for understanding the role of the RKKY interaction on
the ground state order and corresponding transport properties of such
systems, extending beyond the conventional perturbative regime.
\end{abstract}

\maketitle
 
\section{\label{sec:level1} Introduction}
  
Spin-fermion models, consisting of localized degrees of freedom (the
`spins') interacting with itinerant degrees of freedom (the `fermions'),
have proven useful in capturing the essential physics of many correlated
systems in condensed matter physics, including manganites
\cite{mangref}, cuprates \cite{cupref,cupref2}, nickelates
\cite{nref1,nref2}, iron superconductors
\cite{ironscref1,ironscref2,ir3}, heavy fermion materials
\cite{hfref1,hfref2}, and ferromagnetic semiconductors
\cite{fs1,fs2,fs3,fs4}, among others. In these materials, there is often
a natural separation of localized and itinerant electrons, such as the
localized $t_{2g}$ and mobile $e_{g}$ electrons in manganites
\cite{mang-dagotto}, or the localized $3d^{5}$ electrons, which form a
spin $5/2$, and the mobile ${\rm Mo}$ electrons, in the double
perovskite ${\rm Sr_{2}FeMoO_{6}}$ \cite{dpref1,dpref2,dpref3}. In many
such cases, the former have large local moments, and may be approximated
as classical spins, resulting in a simplified model of itinerant
electrons coupled to classical Heisenberg spins. The simplest example of
this is the double exchange model \cite{dex1,dex2}, which has been
applied to study the phenomenon of colossal magnetoresistance
\cite{mang-dagotto} and, under different contexts, the emergence of
Majorana edge states for magnetic chains in contact with an $s$-wave
superconductor \cite{tpsuc1,tpsuc2,tpsuc3,tpsuc4}.

An important ingredient driving the physics in such models is the
effective Ruderman-Kittel-Kasuya-Yosida (RKKY) interaction
\cite{rkky1,rkky2,rkky3} between the spins mediated by the mobile
conduction electrons \footnote{In this paper, we use the term RKKY to
denote the conduction electron mediated spin-spin interaction in all
parameter regimes, and not just in the perturbative limit.}. When an
isolated impurity is inserted in a sea of electrons, it polarizes the
electron sea around it, which can act on another impurity, resulting in
an indirect long range spin-spin interaction that decays as $1/r^{d}$ in
$d$ dimensions, and oscillates in sign with a period determined by the
Fermi vector $k_{F}$. In a lattice of spins, the resultant interactions
lead to a variety of spiral (SP), ferromagnetic (FM) and
antiferromagnetic (AF) phases, generating a very rich phase diagram
\cite{hu-spiral,loss-spiral1,loss-spiral2,maxim-spiral}. 
 
An intriguing situation is created when the spins are embedded in a
superlattice structure with spinless `buffer' regions separating
`blocks' of spins (See Fig.~\ref{fig:Fig_1}). This geometry is realized,
for example, in the classic giant magnetoresistance (GMR) experiments
consisting of alternating thin layers of FM and non-magnetic (NM)
material sandwiched together \cite{Baibich1988,Binasch1989,gmr3,Nobel}.
In such a setup, the simple two-spin RKKY interaction is generalized to
a more complex effective one between the spin blocks, with the NM buffer
regions playing the role of the spin-spin separation. By adjusting the
length of these buffer regions, the FM `blocks' may be aligned
antiferromagnetically with respect to each other. When a magnetic field
is applied, a huge increase in conductivity is observed, lending the
phenomenon its name. The conventional explanation follows from different
scattering amplitudes for up and down spin electrons, leading to a much
smaller resistance when the magnetic field aligns all moments in the
different blocks in the same direction \cite{gmr3,Nobel}. The detailed
spatial character of the RKKY interaction is crucial in determining the
physics of the system.
 
Several existing papers have generalized the theory of the RKKY interaction to larger magnetic clusters \cite{skomski}, or explored
experimentally the nature of magnetic order for small clusters \cite{symrkky} rather than single magnetic impurities.  The theoretical work has not, however, linked these more complex forms of the conduction-electron-mediated interaction to the transport properties, a link which is crucial to modeling GMR phenomena. Moreover, much of this work has been confined to the perturbative RKKY regime, which fails to capture the subtle interplay of strong spin polarization and finite size effects of the magnetic clusters on the ground state, and especially, the transport properties.

Here, we address this outstanding issue by studying the role of the RKKY interaction on the ground state order and corresponding transport properties in a superlattice system. Using a spiral ansatz for the classical spins, we analyse a spin fermion model and investigate the
possible ground states at $T=0$ in various parameter regimes, going far beyond the standard perturbative RKKY regime. We show that, in general, the RKKY interaction can lead to FM, SP and AF order for different parameter values. As the transverse width is increased, however, the SP phases are suppressed. In this regime, we find that the ground state oscillates between FM and AF order as the buffer length is varied. The Drude weight $D$ shows corresponding oscillations as well,
demonstrating enhanced conductivity in the FM state, in agreement with observations from GMR experiments \cite{parkin1,parkin2}. We summarize these results in a series of phase diagrams in the buffer length $L_{0}$ and Hund's coupling $J_{H}$ plane for different values of the chemical potential $\mu$. Analysis of these oscillations shows that, somewhat
surprisingly, their salient characteristics, such as the dependence of
their periods on $\mu$, can be explained from simple considerations of
standard RKKY theory even when the system is far away from the
perturbative regime. On the other hand, transport properties like the spin resolved Drude weight display unexpected results that cannot be derived from the perturbative picture, underscoring the ability of our calculations to capture effects beyond such conventional methods.
% 
% Analysis of these oscillations shows that, somewhat
% surprisingly, their salient characteristics, such as the dependence of
% their periods on $\mu$, can be explained from simple considerations of
% standard RKKY theory even when the system is far away from the
% perturbative regime. 
%% 
%% , where we find an interesting persistence of antiferromagnetic order
%% at large $L_{0}$ for certain $J_{H}$ values, in stark contrast to the
%% behavior at small $L_{0}$, where the system oscillates periodically
%% between FM and AFM states. 
Finally, we discuss the implications of these results, expand on their
connection to the GMR experiments, and conclude by outlining possible
extensions in future work.

\section{Model and Methods} 

%% In general, modelling these systems, which often consist of several
%% elements combined together in elaborate lattice structures, is an
%% extremely complicated task, involving several orbitals,
%% electron-electron interactions and electron-phonon interactions.
%% Nevertheless, much of the basic physics can be captured by a
%% simplified setup consisting of large localized moments, typically
%% originating from partially filled $d$ or $f$ shells in the transition
%% / magnetic atoms, and mobile electrons originating from the same or
%% other atoms in the material. The electrons, which are coupled to the
%% localized spins, mediate an effective RKKY interaction between them.
%% In addition, there are other spin-spin interactions caused by
%% mechanisms such as superexchange, which may result in different kinds
%% of long range order. We model these localized moments by classical
%% spins $S_{i}$, interacting with the itinerant electrons with a Hund's
%% coupling $J_{H}$

%%------------------------------------------------------------------------
\begin{figure}[t]
\centerline{
\includegraphics[scale=0.12,angle=0]{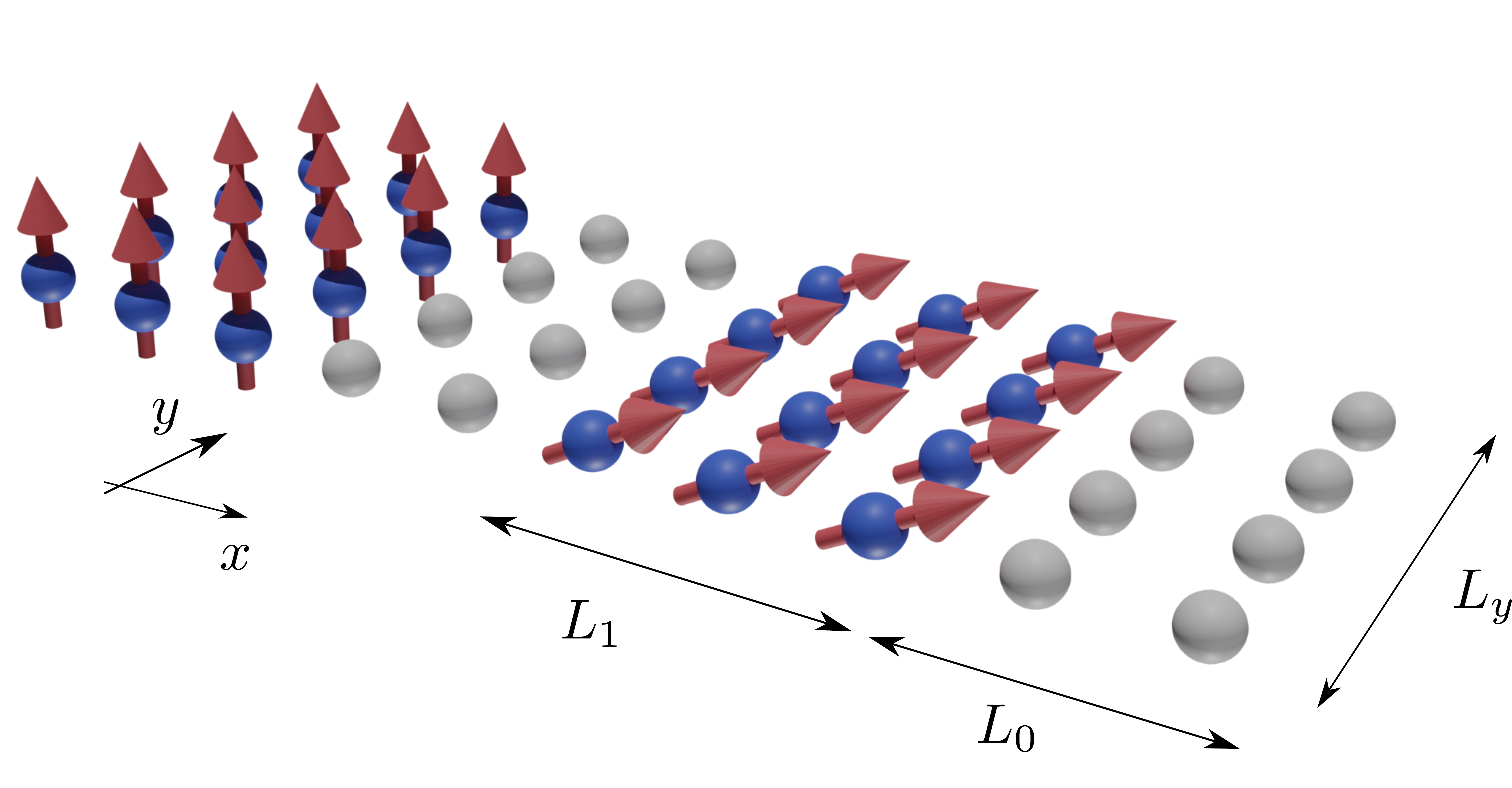}
}
\caption{The superlattice geometry. The blue spheres represent lattice
sites in the spin blocks of length $L_{1}$ possessing classical Heisenberg
spins $\vec{S}_{i}$ (represented by the red vectors), while the
white spheres represent the intermediate buffer sites without spins. The
combined spin and buffer block, of length $L_{c} = L_{0} + L_{1}$, and
width $L_{y}$, is repeated in the x-direction $N_{c}$ times. PBC are
imposed in both the x and y directions. 
%% The lower cartoon depicts the reduced sum of 1D systems the original
%% problem is equivalent to when considering the periodicity in the
%% y-direction, and its associated transverse modes.
}
\label{fig:Fig_1}
\end{figure}
%%-----------------------------------------------------------------------  
 
%% In this work, 
We focus on studying the effect of the RKKY interaction on the spins in a superlattice geometry, inspired by the classic experiments on GMR \cite{Baibich1988,Binasch1989,gmr3,Nobel}. We consider a quasi one-dimensional (1D) lattice with length $L_{x}$ and width $L_{y}$, where the classical spins reside on blocks of length
$L_{1}$, alternating with spinless buffer blocks of length $L_{0}$. This unit of length $L_{c} = L_{0}+L_{1}$ is repeated $N_{c}$ times in the
$x$-direction, and thus $L_{x} = N_{c} L_{c}$, and the total number of lattice points $L_{t} = L_{x}  L_{y}$. We impose periodic boundary
conditions (PBC) along both the $x$ and $y$ directions. Figure~\ref{fig:Fig_1} shows the geometric layout in detail.

In real materials, electron-spin and spin-spin couplings usually originate from a variety of physical phenomena involving multiple orbitals \cite{fazekas}. However, our concern here will be to construct the simplest possible model that describes our system, i.e., FM blocks in a superlattice. Hence, our model parameters should be interpreted as phenomenological, effective variables and not material specific ones.
To begin with, we model the localized moments by classical Heisenberg spins
$\vec{S}_{i}$, which are coupled to the itinerant electrons by a ferromagnetic Hund's coupling $J_{H}$. We include a nearest
neighbour ($\langle \ldots \rangle$) ferromagnetic spin-spin coupling of
strength $J_{F}$ to induce ferromagnetic order inside each spin block, mimicking the ferromagnetic layers in the GMR experiments. The itinerant electrons are characterised by a nearest neighbour hopping $t$ and a chemical potential $\mu$. For simplicity, we assume that the hopping elements are the same in the spin and buffer blocks.
Thus, our Hamiltonian is given by,
\begin{eqnarray}
 {\cal H} &=& (-t)  \sum_{\langle ij \rangle\sigma} \big( c^{\dagger}_{i
\sigma} c^{\phantom{\dagger}}_{j \sigma} + \mbox{h.c.} \big) - \mu
\sum_{i \sigma} c^{\dagger}_{i \sigma} c^{\phantom{\dagger}}_{i \sigma}
\nonumber \\
 && - J_{H} \sum_{i \alpha \beta} \vec{S}^{\phantom{\dagger}}_{i}
\cdot
c^{\dagger}_{i \alpha} 
\vec{\sigma}^{\phantom{\dagger}}_{\alpha \beta}
c^{\phantom{\dagger}}_{i \beta} - J_{F}
\sum_{\langle ij\rangle} \vec{S}_{i}\cdot\vec{S}_{j},
\label{eq:hamil}
\end{eqnarray}
where $c^{\dagger}_{i \sigma}$ ($c^{\phantom{\dagger}}_{i \sigma}$)
denotes the fermionic creation (annihilation) operator at site $i$ with
spin $\sigma$, and $\vec \sigma = (\sigma^x, \sigma^y, \sigma^z)$ is the
vector of spin-1/2 Pauli matrices. The classical spins $S_{i}$ in the second line, as explained before, refer to the localized moments in the ferromagnetic region, and are absent in the buffers. We will further assume that the ratio $J_{F}/ J_{H}$ is large enough so that all spins within a particular block are aligned ferromagnetically at $T=0$, as seen in experiments. We reiterate that the final term, $J_{F} \vec{S}_{i}\cdot\vec{S}_{j}$, is a purely phenomenological intra-block term which acts only on the classical spins $\vec{S}_{i}$ to ensure a totally ferromagnetc orientation inside each block; the interactions between the itinerant electrons and the spins are entirely encapsulated by the Hund's coupling term with strength $J_{H}$. With the above assumptions, we will drop this term from subsequent calculations.
  
Without the $J_{F}$ term, our Hamiltonian is determined by $J_{H}/t$ and $\mu/t$. Keeping $\mu$ constant, the two extreme limits of the model are given by $J_{H}/t = 0$ and $J_{H}/t = \infty$. The former limit corresponds to ferromagnetic blocks that are uncoupled to free electrons on the superlattice. When $J_{H}$ is turned on, a perturbative expansion can be performed, leading to the standard RKKY expression (see Appendix \ref{sec:pert}). The other limit, $J_{H}/t = \infty$, corresponds to a situation where the electrons are localized, leaving individual spins uncorrelated (other than implicitly through the large intra-block $J_{F}$ in our model). In a normal lattice, a small non-zero $t$ leads to the well known spinless double exchange model \cite{sl-de} , with an effective hopping that favours FM order. In our superlattice setup, we also expect a finite $t$ to lead to FM order among the spin blocks, but with a reduced magnitude $\sim t^{2} / J_{H}$ due to the energy mismatch between the electronic states at the spin block boundary as a result of the polarization by $J_{H}$. When $t \sim J_{H}$, the effects of hopping and spin polarization are of the same order of magnitude, and a simplified solution to the problem is no longer possible. 

Our Hamiltonian obeys particle-hole symmetry at $\mu = 0$ (see Appendix \ref{sec:ph_symm}),
which implies that the number density is strictly fixed at $\langle n \rangle = 1$ for all parameter values at this point. Away from this special value, $\langle n \rangle$ could vary in principle with the parameters, especially the buffer length $L_{0}$. In practice, however, we find that $\langle n \rangle$ saturates with $L_{0}$ fairly quickly over a large range of $\mu$ (See Appendix \ref{sec:nden}). Furthermore,
in a typical experimental setup, $\mu$ can be controlled 
%% far more easily
by adjusting the gate voltage. 
%% compared to $\langle n \rangle$. 
In view of this, our calculations will be done at fixed $\mu$ instead of
$\langle n \rangle$.

Since this Hamiltonian is quadratic in the fermions, it can be solved by
diagonalizing a matrix of size $2 L_{t} \times 2 L_{t}$ for any
arbitrary configuration of the spins $\big\{ \vec{S}_{i} \big\}$. 
The doubling occurs due to mixing of the electron spins by the Hund's
coupling term. However, as our spins are distributed on a periodic
superlattice geometry, the ground state at $T=0$ is expected to be
regular, and not random. It is well known that in a regular lattice
geometry, with spins on each site, the competition between the Fermi
wave vector $k_{F}$ and $J_{H}$ leads to a variety of spiral spin
configurations at different fillings in 1D
\cite{hu-spiral,loss-spiral1,loss-spiral2}. Motivated by these results,
we assume a similar spiral ansatz for our quasi 1D superlattice system.
The width $L_{y}$ provides an added ingredient in our system due to the
presence of multiple transverse modes.
%% Even though we restrict ourselves to $\mu=0$, due to the added
%% ingredients including the width $L_{y}$ and the superlattice
%% structure, our model could support non-trivial spiral phases even at
%% half-filling.

Following the reasoning in Refs. \cite{hu-spiral,tpsuc1,tpsuc2}, we argue that in the absence of spin orbit coupling (leading to antisymmetric spin-spin interactions such as the Dzyaloshinskii-Moriya coupling \cite{dm1,dm2}), at zero external magnetic field, a planar ground state ansatz results, in no loss of generality. Hence,
we assume a spiral state of the following form:
\begin{eqnarray}
\label{eq:ansatz}
 {\cal S}^{m}_{x} &=& {\rm cos} (q x_{m}) \nonumber \\{\cal S}^{m}_{y} &=& {\rm sin} (q x_{m})  \\
 {\cal S}^{m}_{z} &=& 0. \nonumber
\end{eqnarray}
Here, ${\cal S}^{m}$ denotes any of the local spins in the `block' $m$,
$x_{m}$ denotes the `block' coordinate, and $q = 2 \pi n / N_{c}$
($n=0,\ldots,N_c-1$), denotes a `block' momentum index. This specific
ansatz has been chosen due to convenience; any configuration resulting
from a global rotation of the above state would be equally valid, due to the preserved SU(2) symmetry of the Hamiltonian. 

With this ansatz, the solution of the Hamiltonian can be reduced to that
of a 1D block of length $L_{c}$ for each value of two Fourier
coefficients $k_{y}$ and $P$, to be described in detail below. To derive this, we first write the original
Hamiltonian by redefining the lattice index $i \equiv (a,m,y)$, where
$a$ denotes the $x$-coordinate measured from the first site of the same
`block', $m$, the block number, and $y$, the coordinate in the $y$
direction. With these definitions, we can expand the fermionic annihilation operators:
%%  in the following manner:
\begin{eqnarray}
 c_{i \sigma} \equiv c_{a,m,y,\sigma} &=& \sum_{k_{y},P} c_{a,k_{y},P,\sigma}
e^{{\rm i} ( k_{y} y + P x_{m} )}.
 \label{eq:ft}
\end{eqnarray}
Here, $k_{y} = 2 \pi n / L_{y}$ is the usual Fourier coefficient in the $y$
direction, and $P = 2 \pi m / N_{c}$ is a `block' Fourier coefficient in
the $x$ direction, similar to $q$ in the spiral ansatz. 

Using these definitions, the Hamiltonian can be recast: 
%% in the following form:
\begin{eqnarray}
 {\cal H} &=&  \sum_{k_{y}, P, \sigma} \Big\{ \sum^{L_{c} - 1}_{a = 1} (-t) ~ \big( c^{\dagger}_{a+1, k_{y}, P, \sigma} c^{\phantom{\dagger}}_{a, k_{y}, P, \sigma} + {\rm h.c.}  \big)  \nonumber \\ 
 &&+ ~~(-t) ~\big( c^{\dagger}_{1, k_{y}, P, \sigma} c^{\phantom{\dagger}}_{L_{c}, k_{y}, P, \sigma} e^{- i P} + {\rm h.c.} \big) \nonumber \\ 
 && -~ \sum_{a}~\big( 2t {\rm cos} ~(k_{y}) + \mu \big) ~c^{\dagger}_{a,k_{y}, P, \sigma} c^{\phantom{\dagger}}_{a, k_{y}, P, \sigma} \Big\} \nonumber \\ 
 && - J_{H} \sum_{a,k_{y},P} \big\{ c^{\dagger}_{a,k_{y},P,\uparrow} c^{\phantom{\dagger}}_{a,k_{y},P+q,\downarrow} + {\rm h.c.} \big\}.
 \label{eq:hamil_k}
\end{eqnarray}

\begin{figure}[t]
\centerline{
\includegraphics[scale=0.32]{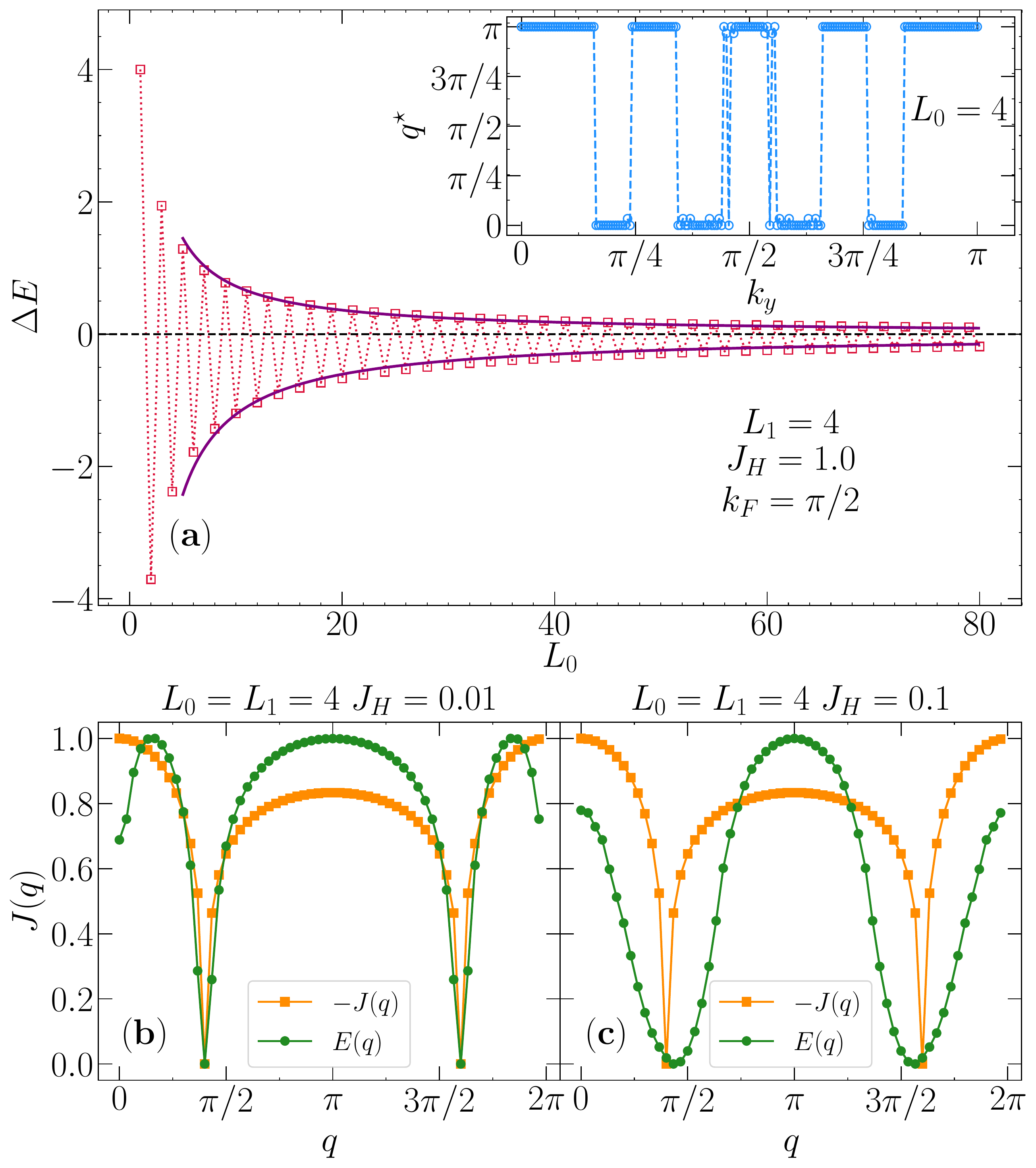}
}
\caption{(a) Energy difference $\Delta E$ between AF and FM states with increasing $L_{0}$, at $k_{F}=\pi/2$; $L_{1}=4$ and $J_{H} = 1$. The period of oscillation is $2$, and the envelope goes as $1/L_{0}$, consistent with the RKKY form $\sim {\rm cos}(2 k_{F} r) / r = {\rm cos}(\pi L_0) / L_0$. Inset shows that the optimum $q = q^{\star}$ value flips between FM and AF with changing chemical potential $\mu$ for the same parameters, fixing $L_0=4$. (b) and (c) show the dispersion $E(q)$ (green circles, scaled) for $J_{H}=0.01$ and $0.1$ at $k_{F} = \pi / 10$.  SP order at small $J_{H}$ gradually evolves towards FM/AF with increasing $J_{H}$. Orange squares show results from second order perturbation for small $J_{H}$ (also scaled). In second order, $\delta E_{2}(q) = - \sum_{m,m'} J_{m,m'} \vec{S}_{m}(q).\vec{S}_{m'}(q) = - {\rm Re}( J(q) )$ (see Appendix~\ref{sec:pert} for details), whose minimum coincides with that of $- {\rm Re}(J(q))$, as the plots demonstrate.}
% 
% Here, $\delta E = - \sum_{i,j}
% J_{ij} \vec{S}_{i}.\vec{S}_{j} = {\rm const.} - \sum_{q} J_{q} |
% S(q)|^{2}$. Since $| S(q)|^{2}$ is positive, the optimum $q=q^{\star}$
% is given by the minimum of $-J(q)$. Orange lines plot the scaled $-J(q)$
% (see Appendix~\ref{sec:pert} for details). 
% }
\label{fig:1d}
\end{figure} 
%-----------------------------------------------------------------------

For any Fourier mode $(k_{y},P)$, the above Hamiltonian is equivalent to a
1D tight-binding Hamiltonian of a block of length $L_{c}$, at
an effective chemical potential $\mu(k_{y}) = -2 t~ {\rm cos} (k_{y}) - \mu$.
The Hund's term mixes the $P,\uparrow$ state for each site $a$
with the corresponding $P+q,\downarrow$ state. 

Each solution provides $2 L_{c}$ eigenvalues $\lambda^{q}_{k_{y},P,n}$,
%% where $n \in [0, 2L_{c})$, and the total free energy at $T=0$ is given
where $n \in [0, 2L_{c}]$, and the ground state energy is given
by  $E(q) = \sum_{k_{y},P,n} \lambda^{q}_{k_{y},P,n} \big(1 -  \theta(
\lambda^{q}_{k_{y},P,n}) \big)$, where $\theta(x)$ is the usual Heaviside
theta function. To find the optimum value $q = q^{\star}$ for any given
set of parameters, we calculate $E(q)$ for all allowed values of $q$ and
find the minimum. We solve these equations for various values of $\mu$,
$J_{H}$, $L_{0}$, $L_{1}$ and $L_{y}$ to gain insight into the effect of
the RKKY interaction on the ground state order.

To investigate the transport properties, and explore parallels with the
GMR systems, we calculate the Drude weight, which is an indicator of the
DC conductivity. In linear response, the Drude weight $D$
is determined by the current-current correlation function $\Lambda_{xx}$
by \cite{sf-form}  
\begin{eqnarray}
  \frac{D}{\pi e^{2}}&=&\langle -k_{x} \rangle -{\rm Re}~ \Lambda_{xx}(q=0,\omega \rightarrow 0).
\end{eqnarray}
Here, $\langle -k_{x} \rangle$ denotes the kinetic energy along the x
direction in the system, and $\Lambda_{xx}(q=0,\omega \rightarrow 0)$ is
the Fourier transform of the current-current correlation function
$\Lambda_{xx}(i-j,\tau-\tau^\prime) \propto\langle J_{x}(i,\tau)
J_{x}(j,\tau^\prime) \rangle$, where $J_{x}$ is the current operator in
the x-direction given by $ J_{x} = (-{\rm i} t) \sum_{i,\sigma} \big(
c^{\dagger}_{i+x, \sigma} c^{\phantom{\dagger}}_{i, \sigma} - {\rm h.c.}
\big)$. The current-current correlation function can be calculated in a
straightforward manner since it is quadratic in the
fermions, and further simplified using the Fourier components $(k,P)$ as
in the case of the Hamiltonian (see Appendix \ref{sec:cond} for
details). This provides us with the crucial tool to explicitly calculate
and connect the transport with the ground state properties, constituting
an advance over previous theoretical work
~\cite{Edwards91}.
%% on such or related systems~\cite{Edwards91}.

\section{\label{sec:level2} Results}

We now present results on the ground states for various parameter
values.
%%  by solving the equations previously presented. 
As seen in Eq.~\ref{eq:hamil_k},
%% can reduce the Hamiltonian to a set of 1D tight binding Hamiltonians
%% with an effective chemical potential $\mu(k)$ corresponding to each
%% transverse mode $k$. Thus, 
the system is effectively a sum of 1D
superlattices with an effective chemical potential $\mu(k_{y})$
corresponding to each transverse mode $k_{y}$. 
%% Due to the presence of spiral phases in 1D systems discussed before, 
%% our system may have spiral ground states even at half-filling.  
To gain more insight,
%% into our problem, 
we start by considering the 1D case alone. This will form the
basis for understanding the case of wide superlattices. In
what follows, the energy scale is set by $t = 1$. A majority of the results will be presented for $J_{H} = 1$. Using detailed phase diagrams, we will later demonstrate that the system properties are remarkably similar over a large range of $J_{H}$ values about this point.

\subsection{1D superlattices}

Figure~\ref{fig:1d} summarizes our results on one-dimensional
superlattices for different parameter values. We start by demonstrating
the effect of the spatially varying RKKY interaction in our system in
Fig.~\ref{fig:1d}(a), by showing how the lowest energy state oscillates
between FM and AF configurations with changing buffer length $L_{0}$ at
half filling [$\mu(k_{y}) = 0$; we denote $\mu$ as a function of
`$k_{y}$' to remind us of the connection with our original quasi-1D
system], in a typical superlattice with $L_1=4$. If we define a nominal
$k_{F}$ by $-2t~ {\rm cos} (k_{F}) = -\mu(k_{y})$, then we have $k_{F} =
k_{y} = \pi/2$. The period of oscillations and the envelopes of the
$\Delta E \equiv E_{\rm FM} - E_{\rm AF}$ curve are in agreement with
the RKKY form $\sim {\rm cos}(2 k_{F} r) /r$. Here, as explained
earlier, the distance $r$ between the magnetic `impurities' of the
original interaction is set by the buffer size $L_0$ separating the
$L_1$-sized magnetic regions. The inset shows the optimum values
$q^{\star}$ for different $\mu$ at $J_{H}=1$ and $L_0=4$. We find that
at this value of $J_{H}$, the ground state is FM or AF for most of the
$\mu$ values, and SP phases are largely suppressed.

On the other hand, Figs.~\ref{fig:1d}(b) and \ref{fig:1d}(c) show the
dispersion relations $E(q)$ for smaller values of $J_{H} = 0.01$ and
$0.1$, respectively, for $k_{F} = \pi / 10$. As an independent check on our calculations, we
also compare these with the results from a second order perturbation
expansion for small $J_{H}$ (see figure caption and
Appendix~\ref{sec:pert} for details). We find that at $J_{H}=0.01$, the
optimum $q$ is spiral, and as expected, perturbation theory works well.
As $J_{H}$ is increased, the optimum $q$ tends to move to either FM or
AF, and the agreement with perturbation results also becomes worse, as
one would expect, shown by the results for $J_{H} = 0.1$.

\subsection{Quasi-1D superlattices}

%------------------------------------------------------------------------
\begin{figure}[t]
\centerline{
\includegraphics[scale=0.32]{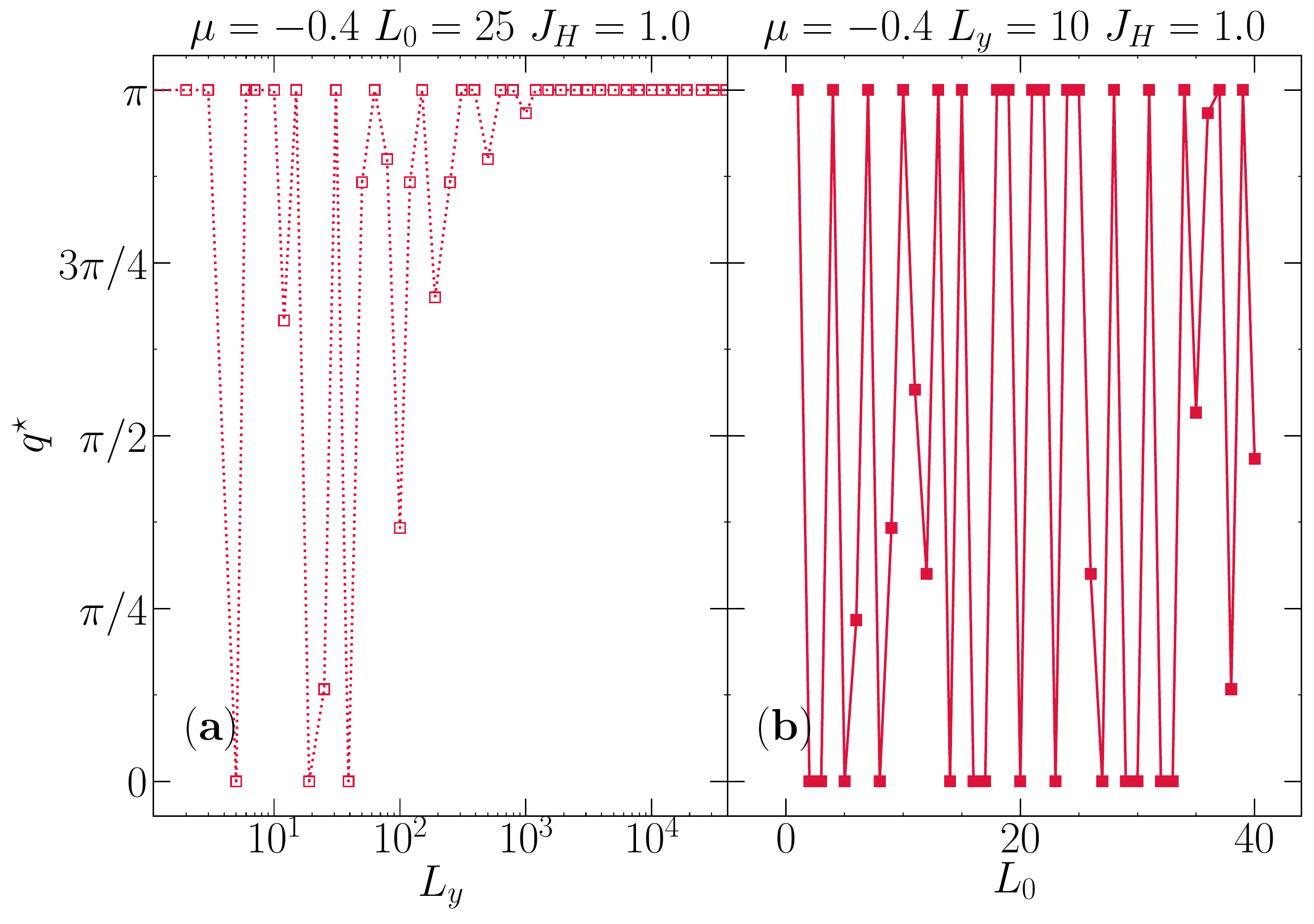}
}
\caption{(a) Optimum value $q^{\star}$ over 
%% a large range of 
$L_{y}$ values $\sim [10, 10^{4}]$ at $\mu = -0.4$, $L_{0}= 25$, $L_{1} = 4$ and
$J_{H} = 1.0$. We find that in a large range $L_{y} \lesssim
10^{3}$, the ground state consists of various SP orders, whereas for
large $L_{y} \gtrsim 10^{3}$ these are suppressed and the system has an
AF ground state at the chosen parameter values. (b) shows optimum $q =
q^{\star}$ vs $L_{0}$ for small $L_{y}=10$, where we see rapid
oscillations without any uniform periodicity (see text for more
details). The number of cells $N_c$ is kept at 60.}
\label{fig:fig_3}
\end{figure}
%-----------------------------------------------------------------------

In this subsection, we present our main results on the quasi-1D
superlattice system
%% . As discussed earlier, this can be thought of as a
which reduces to a
collection of separate 1D systems at a range of chemical potentials
$\mu(k_{y}) = -2t~{\rm cos}(k_{y}) - \mu$, defined by the normal modes
$k_{y} = 2 \pi n / L_{y}$, with $n$ varying from $0$ to $L_{y}-1$. We
start by probing the effect of the width $L_{y}$ on the system
properties. Then, we  move on to the transport properties by
calculating $D$ explicitly and analysing its correlation with the ground
state spin configuration. 
We present detailed phase diagrams to study the ground state
dependence on $\mu$ and $J_{H}$, noting how its periodicity changes with
$\mu$. Finally, we discuss this periodicity and its dependence on $\mu$
and other system parameters in detail, and provide a simple RKKY
framework that, somewhat surprisingly, explains a majority of these observations.
% provide a simple RKKY
% framework that explains a majority of these observations.

\subsubsection{Dependence on $L_{y}$}

To relate our results to the GMR experiments better, where the
transverse dimensions are orders of magnitude larger than the lengths
$L_{0}$ and $L_{1}$, we explore our system over a large range of values
of $L_{y} \in [{\cal O}(10) , {\cal O}(10^{4}) ]$, while keeping $L_{0},
L_{1} \sim {\cal O}(10)$. Figure~\ref{fig:fig_3} summarizes these
results in a nutshell. Figure~\ref{fig:fig_3}(a) shows the optimum
$q=q^{\star}$ for different values of $L_{y}$ up to $10^{4}$, for fixed
$\mu$, $L_{0}, L_{1}$ and $J_{H}$. We find that over a large range
$L_{y} \lesssim 10^{3}$, the system consists of a variety of SP orders
until $L_{y} \gtrsim {\cal O}(10^{3})$, where they are suppressed and
the ground state displays AF order at our parameter values. We emphasize
that this is the result of a sum over all transverse modes $k$, and
depends intricately on the energy balance of the various orders at each
$\mu(k)$ [see Eq.\eqref{eq:hamil_k}].

In comparison, we further report in Fig.~\ref{fig:fig_3}(b) a
representative plot of the optimal $q = q^{\star}$ for a small fixed
$L_{y} \sim {\cal O}(10)$ but with varying buffer size $L_{0}$, at the
same $\mu$ and $J_{H}$ values.  We find that the system shows rapid
oscillations between FM and AF with intermediate SP states, but there is
no clear periodicity in their variation. We will discuss this effect and
contrast it with the behavior at large $L_{y}$ in detail below.
%% An intriguing observation is that the period of oscillation is
%% $L_{y}/2$, as we have checked using other superlattice configurations
%% in this range. A crude explanation of this follows from noting that
%% the Hamiltonian is a sum of 1D Hamiltonians for different $k$ modes
%% associated with the periodicity in $y$. For each mode, the effective
%% chemical potential of the 1D system is given by $\mu_{{\rm eff}} =
%% -2t~{\rm cos}(k)$, and this energy scale controls the nature of the
%% RKKY interaction in that model, including the oscillation period,
%% since it determines the spatial properties of the eigenstates
%% $\psi(\mu_{{\rm eff}})$ of the 1D model around that energy. The
%% overall oscillation period of the total Hamiltonian would be
%% determined by the smallest $k$ mode, given by $k_{{\rm min}} = 2 \pi /
%% L_{y}$. Now, if we assume that $\psi(\mu_{{\rm eff}})$ have similar
%% spatial characteristics to the mode $k$ by naively identifying $k_{F}
%% \simeq k$, then the periodicity immediately follows from the standard
%% RKKY form $\sim {\rm cos}(2 k_{F}r) / r$.

%------------------------------------------------------------------------
\begin{figure}[t]
\centerline{
\includegraphics[scale=0.32]{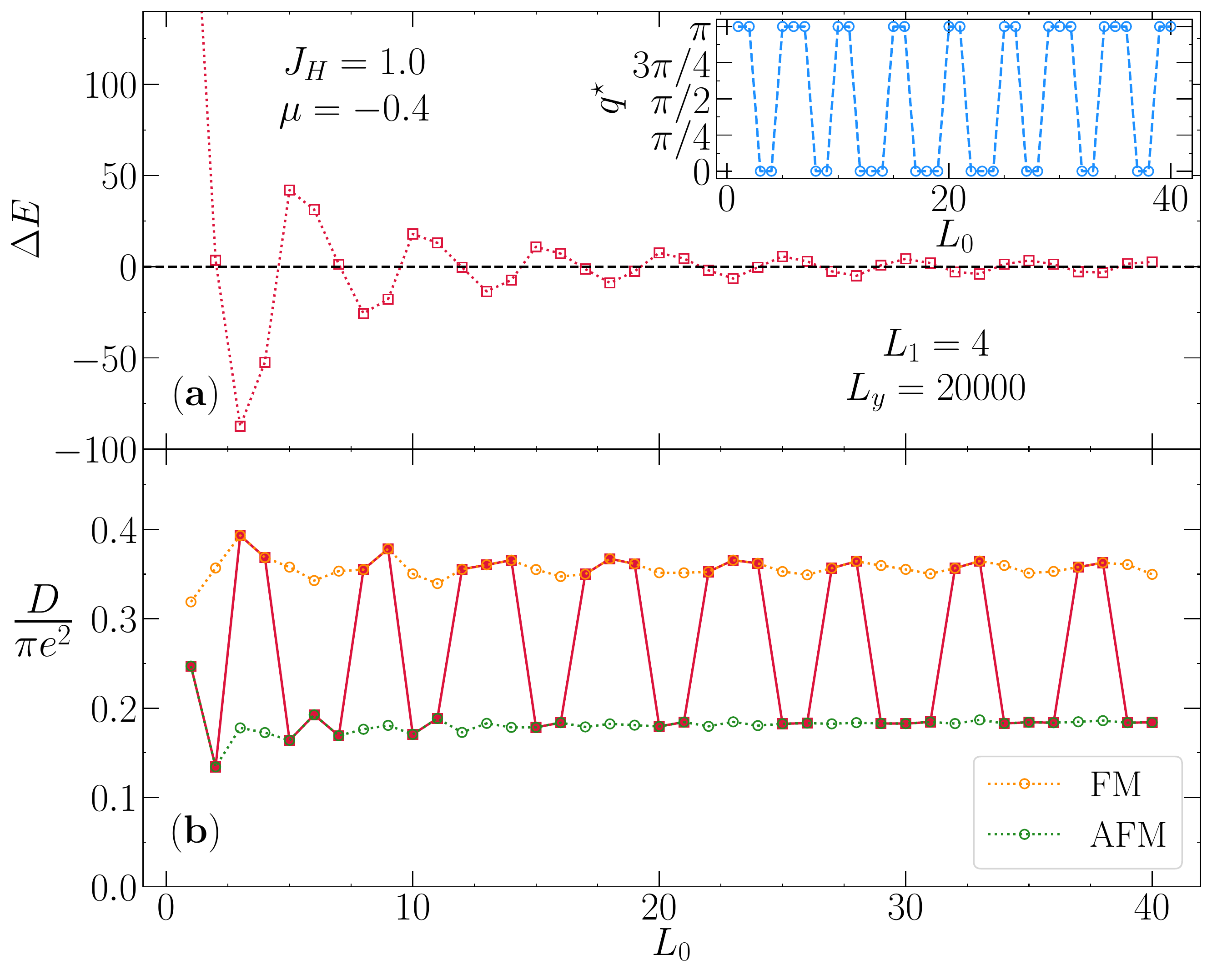}
}
\caption{(a) Energy difference $\Delta E$ beween FM and AF states (main)
and optimum value $q^{\star}$ (inset) vs $L_{0}$ for very wide
superlattices ($L_{y} = 20000$) at $\mu = -0.4$, $J_{H} = 1$ and $L_{1}
= 4$. SP phases are suppressed in this regime, but in contrast to the
results for $L_{y} \sim {\cal O}(10)$, the ground state oscillates
between FM and AF with a period $\sim$ 5 sites. $\Delta E = E_{\rm
FM}-E_{\rm AF}$ shows corresponding oscillations with a magnitude that
decays with $L_{0}$. (b)  the Drude weight $D$ for the FM (orange) and
AF (green) states for the same parameters. The FM state value is $\sim
2-3$ times larger throughout, leading to large oscillations in $D$ (red)
as the ground state keeps flipping between FM and AF. The number of
cells $N_c$ is kept at 60.}
\label{fig:fig_4}
\end{figure} 
%-----------------------------------------------------------------------

\subsubsection{Transport properties}

Having demonstrated that the SP phases are suppressed when $L_{y}\gg
L_{0},L_{1}$, we consider a specific case with $L_{y} = 20000$, in order
to investigate the influence of the magnetic ordering of the classical
spins on the all-important transport properties of the fermions.
Figure~\ref{fig:fig_4} summarizes our results on the ground state and
the Drude weight for this case, at $\mu = -0.4$ and $J_{H} = 1$. The
inset in Fig.~\ref{fig:fig_4}(a) shows that in contrast to the behavior
at small $L_{y}$, the system at large $L_{y}$ oscillates periodically
between FM and AF states with a period $\sim 5$ lattice sites.  In the
main plot, Fig.~\ref{fig:fig_4} (a), the energy difference $\Delta E =
E_{\rm FM} - E_{\rm  AF}$ is plotted against $L_{0}$. The curve
oscillates between positive and negative values as the ground state
changes from AF to FM respectively. The magnitude of the oscillation
decays with increasing $L_{0}$, in keeping with the intuitive
expectation that the RKKY interaction should fall off with increasing
distance between the magnetic blocks.

Figure \ref{fig:fig_4} (b) shows the corresponding results for the Drude
weight $D$ for both the FM and AF states. We find that the conductivity
in the FM state is consistently $2-3$ times higher than the AF state
throughout the whole $L_{0}$ range. As a result, the actual $D$ shows
robust, large oscillations in response to the periodic shifts in the
ground state order. The amplitude of the oscillations appears to remain
approximately constant throughout the whole $L_{0}$ range in the plot.
However, as we will elaborate in the Discussion section
(Sec.~\ref{sec:disc}) below, this is an artifact of being confined to
$T=0$. At any finite temperature, we should expect a gradual suppression
with increasing $L_{0}$.

\subsubsection{Phase diagrams}

%%------------------------------------------------------------------------
\begin{figure}[t]
\centerline{
\includegraphics[scale=0.37]{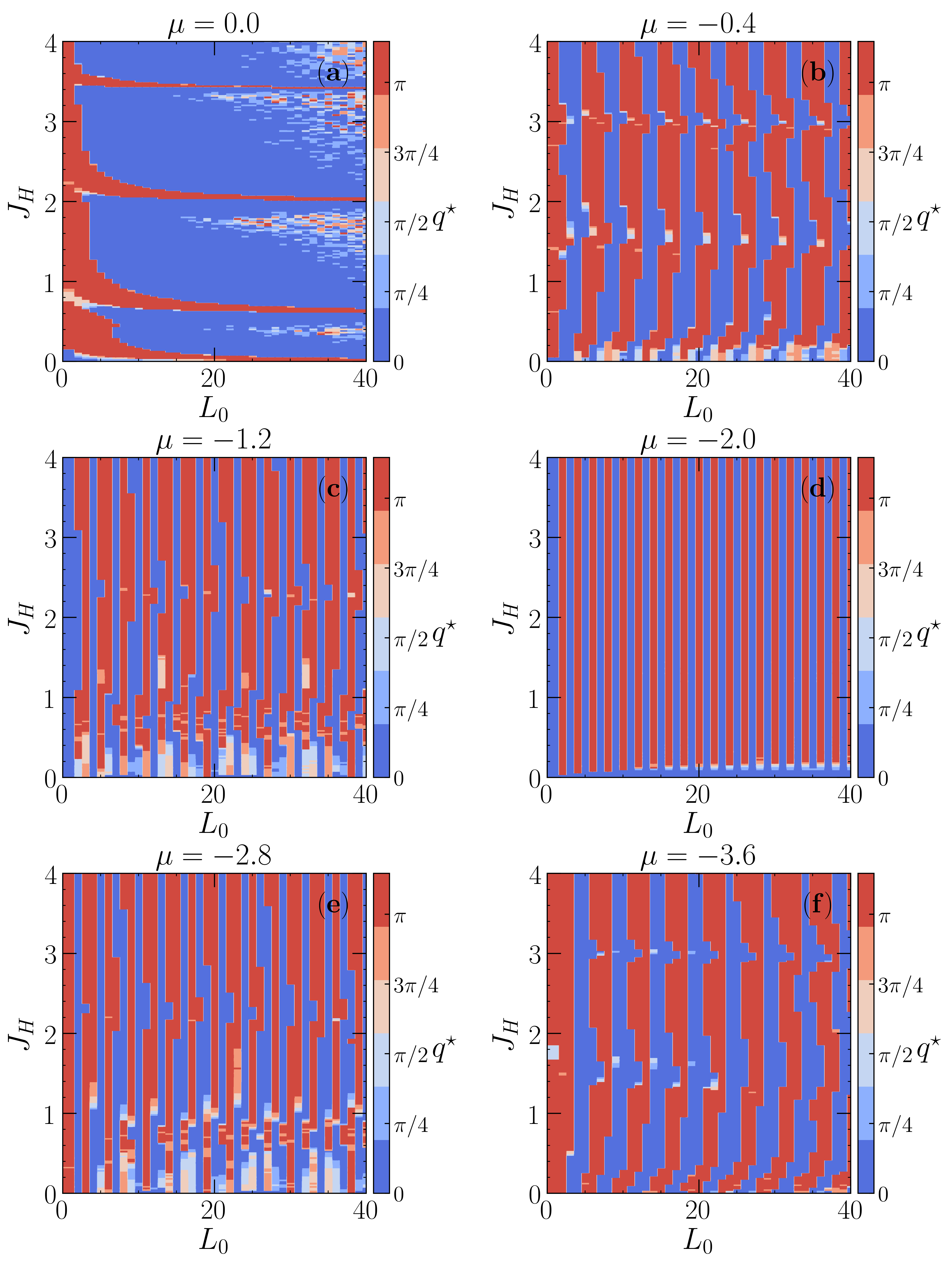} } \caption{Phase diagrams in
$L_{0}-J_{H}$ plane for different $\mu$ values, at $L_{y} = 20000$ and $L_{1} = 4$. Away from half-filling
($\mu \neq 0$), the systems show periodic oscillations between FM and AF
with $L_{0}$ over essentially the whole range of $J_{H} \lesssim 4$. The
basic period is approximately constant for a given $\mu$, but becomes
smaller as it is reduced, reaching a minimum of $2$ sites at $\mu = -2$,
beyond which it increases again, displaying an approximate symmetry
around this value. In contrast, $\mu=0$ (and $\mu=-4$, not shown) show
no periodicity, and persistent AF (FM for $\mu = -4$) regions forming
narrow `fingers' running across the plots (see text).
}
\label{fig:fig_5}
\end{figure} 
%-----------------------------------------------------------------------

Having analysed the ground state and transport properties for a specific
value of $\mu$ and $J_{H}$, we would like to have more insight into how
the system evolves as we vary these parameters. To that end, we
construct detailed phase diagrams in the $L_{0}-J_{H}$ plane with
varying $\mu$, at large $L_{y} =20000$ and $L_{1} = 4$, as shown in Fig.~\ref{fig:fig_5}. We find that away from
half-filling ($\mu \neq 0$), the ground state shows periodic
oscillations with $L_{0}$ for all values of $\mu$. While the pattern
changes somewhat with increasing $J_{H}$, the basic periodicity remains
surprisingly constant over the whole range $J_{H} \lesssim 4$.

On the other hand, the periodicity shows a strong and systematic
dependence on $\mu$, reducing monotonically as $|\mu|$ is increased
until it reaches a minimum of 2 sites at $\mu = -2$, reminiscent of the
results at half-filling for 1D systems, and then increasing again. The
results show an intriguing approximate (but not exact) symmetry around
this value, as evidenced by the similarity of the periods and patterns
at, for example, $\mu = -0.4$, and its counterpart, $\mu = -3.6$. By the
time the system is close to half-filling, $\mu = 0$, the period is so
large that it is not discernible at the relevant length scales $L_{0}
\lesssim 40$. Instead, we find that the phase diagram is mostly
dominated by the FM phase at moderate to large $L_{0}$, while the AF
phase dominates at small $L_{0}$. Remarkably, at certain values of
$J_{H}$, the AF phase extends across the full $L_{0}$ range, forming
narrow fingers running across the plot. At $\mu = -4$ (not shown), the
roles are reversed, and it is the FM phase that forms narrow fingers
across the plot.

%% We find that, in general, the AF phase dominates at small $L_{0}$,
%% while the FM phase is ubiquitous at moderate buffer sizes. At large
%% $L_{0}$, we also find a variety of SP phases, in addition to the FM
%% one. Intuitively, the two extremes may be understood from the fact
%% that at small $L_{0}$, the system is very densely packed with spins,
%% where AF order is preferred as long as $J_{H}$ is not too small. On
%% the other hand, at large $L_{0}$, the system consists of spin blocks
%% very far away from each other, and their interactions are weak, so all
%% possible states are very close in energy. 

\subsubsection{Analysis of periodicity}

%------------------------------------------------------------------------
\begin{figure}[b] 
\centerline{
\includegraphics[scale=0.36]{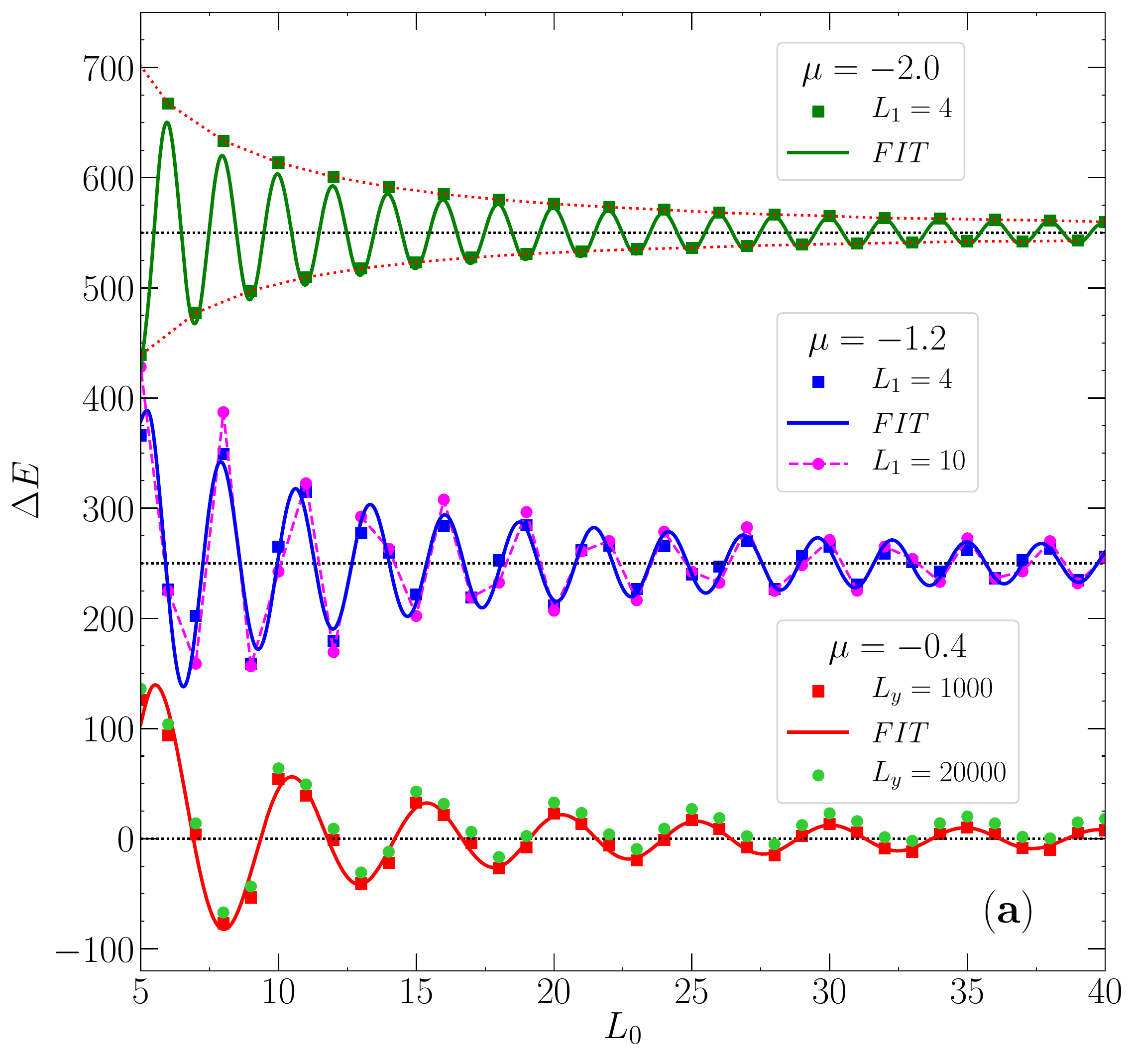}
}
\centerline{ 
\includegraphics[scale=0.36]{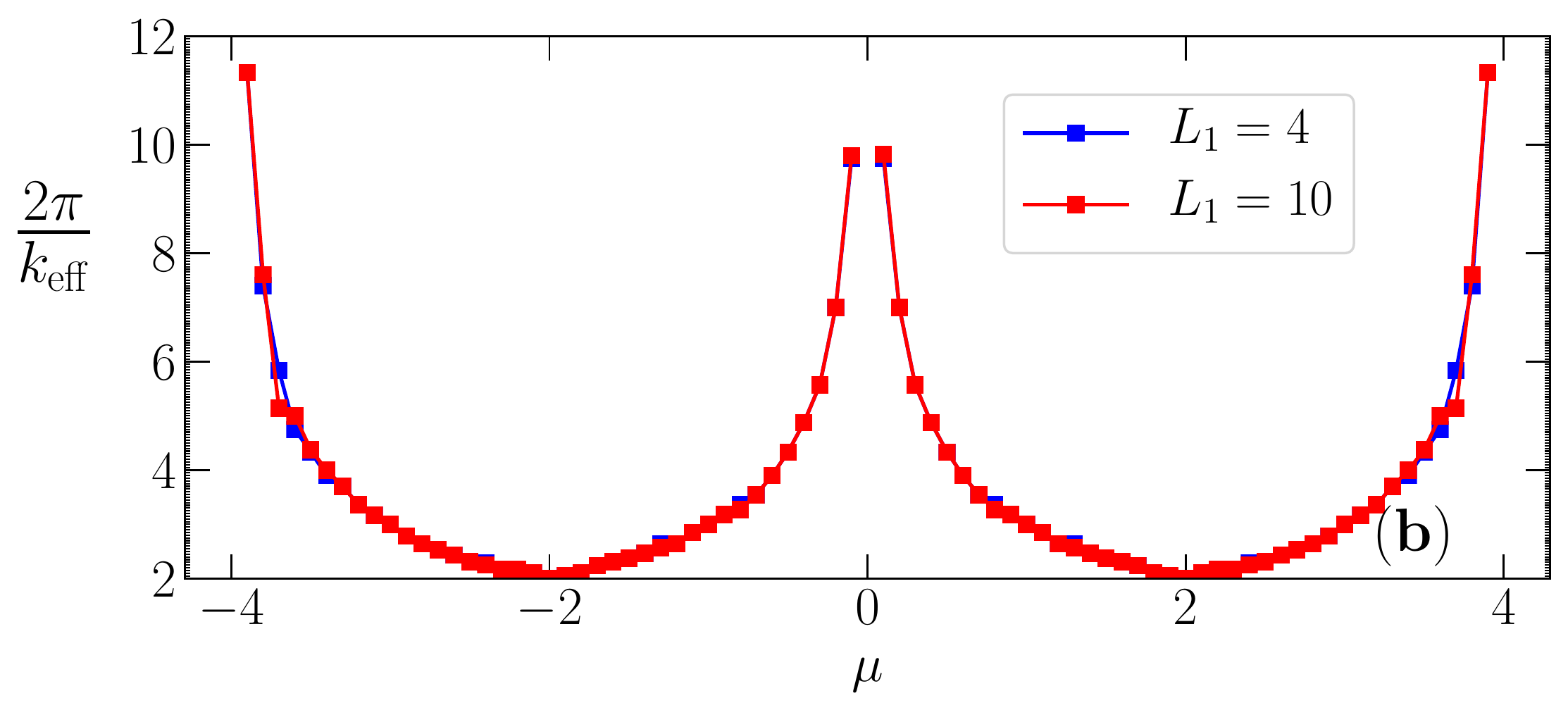}
}
\caption{(a) shows the $\Delta E$ vs $L_{0}$ plots (origins shifted for
clarity, denoted by dotted black lines for each curve) at $\mu = -0.4$
(red squares), $-1.2$ (blue squares) and $-2.0$ (green squares). The
solid lines of same colour show fits to the form $f(L_{0}) \sim a~{\rm
cos}(k L_{0} + b) / L_0^{l}$. Plots of $L_{y} = 20000$ at $\mu = -0.4$
(green circles, scaled by factor of $20$, shifted slightly for clarity)
and $L_{1} = 10$ at $\mu = -1.2$ (magenta circles) show the that the
basic period is independent of these parameters. Red dotted curves at
$\mu = -2$ highlight the decaying amplitude of oscillations with
increasing $L_{0}$. (b) shows the period (expressed as $2 \pi / k_{{\rm
eff}}$) extracted from the fits as a function of $\mu$. The number of
cells $N_c = 60$.
}
\label{fig:fig_6}
\end{figure} 
%-----------------------------------------------------------------------

The results on the periodicity of the ground state and transport deserve
further analysis. In Fig.~\ref{fig:fig_6}(a), we summarize all the
relevant information about the ground state order at large $L_{y}$ and
its variation with $L_{0}$ at different values of $\mu$, $L_{1}$ and
$L_{y}$ (we have already established the remarkable consistency of these
results over a large range of $J_{H}$ values). 

To begin with, we show plots of $\Delta E$ vs $L_{0}$ for three values
of the chemical potential $\mu = -0.4$, $-1.2$ and $-2.0$ respectively,
at $L_{1} = 4$ and $L_{y} = 1000$. To extract the average period of
oscillations, we fit these curves to a functional form $f(x) \sim a~{\rm
cos}(kx + b) / x^{l}$ (similar to the perturbative RKKY form), where
$a$, $b$, $k$ and $l$ are fitting parameters. The resultant fits are
shown by the solid lines in the plot and demonstrate that even though
the oscillations are, in general, complex and not simple cosines by any
means (see the $\mu = -1.2$ plot for example), the basic periodicity, as
well as the decaying envelope of the amplitude, can be fit remarkably
well by such a simple function.

% and demonstrate that even though
% the oscillations are, in general, complex and not simple cosines by any
% means (see the $\mu = -1.2$ plot for example), the basic periodicity, as
% well as the decaying envelope of the amplitude, can be fit remarkably
% well by such a simple function.

In addition, to show the dependence on $L_{y}$ and $L_{1}$, we include
plots of $L_{y} = 20000$ (appropriately scaled by a factor of $20$) and
$L_{1} = 10$ at $\mu = -0.4$ and $-1.2$ respectively.  The former
plot is almost identical to $L_{y} = 1000$ (the circles have been
shifted slightly in the plot for visual clarity), demonstrating that
beyond $L_{y} \gtrsim 1000$,  there is no change in the system
characteristics with further increase in $L_{y}$, beyond a trivial
scaling of the energies. The $L_{1} = 10$ plot, on the other hand, shows the same periodicity but a different `structure' compared to $L_{1} = 4$.
This establishes that the finite size of the spin blocks is responsible,
at least in part, for the detailed internal structure of the curves but
the basic periodicity does not seem to depend on these details.

In Fig.~\ref{fig:fig_6} (b), the period extracted from the fitting
functions (expressed in terms of an effective wavenumber, $k_{{\rm
eff}}$) is plotted against the chemical potential $\mu$ for two
different $L_{1}$ values. As discussed above, the basic periodicity is
independent of $L_{1}$, and shows a monotonic decrease from $\mu = 0$ to
$\mu = -2$, after which it rises symmetrically towards $\mu = -4$. Due
to the particle-hole symmetry at $\mu = 0$, the plot is, of course,
symmetric about half filling.

\begin{figure}[t] 
\centerline{
\includegraphics[scale=0.36]{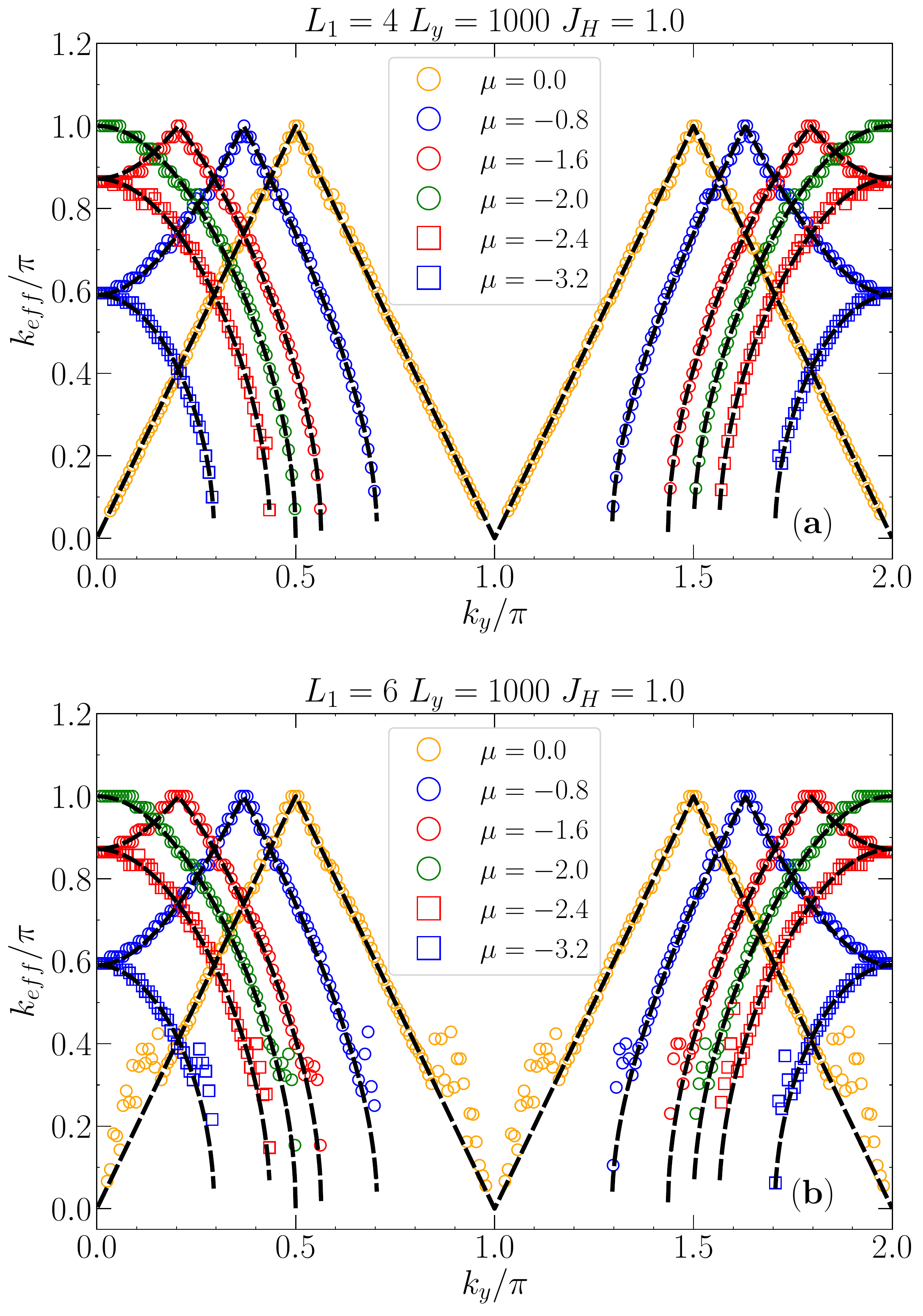}
}  
\caption{The fitting wavevector $k_{{\rm eff}}$ as a function of $k_{y}$
for $L_{1}= 4$ and $6$, $L_{y} = 1000$ and $J_{H} = 1$ for different
values of $\mu$.  `Conjugate' pairs, symmetrically located around $\mu =
-2$ (see text), have the same colour but different symbols (circles and
squares). Black lines show $2k_{F}$, where $k_{F}$ is defined by the
usual relation $k_{F} = {\rm cos^{-1}}[  {\rm cos}(k_{y}) + \mu/2]$. The
close fit shows that the RKKY form using the standard Fermi vector
$k_{F}$ for the buffer regions is valid for $\Delta E = E_{\rm FM} -
E_{\rm AF}$ even outside the perturbative regime for $J_{H}$. The slight
deviations for $L_{1} = 6$ stem from the fitting function algorithm
(fitting errors are not plotted for clarity), and is not indicative of a
sudden change in the periodicity of the system. Effective period of the
quasi-1D system seems to be dominated by the $k_{y}=0$ value.
}
\label{fig:fig_7}
\end{figure}

In order to explain the periodicity, we take a cue from our previous
analysis and repeat the fitting process for each independent transverse
mode $k_{y}$, using the same form as before to extract the basic period.
The result is shown in Fig.~\ref{fig:fig_7}, where we plot $k_{{\rm
eff}}$ as a function of $k_{y}$, for two values of $L_1$. The black
dashed lines plot $2 k_{F}$, defined by $-2t~{\rm cos}(k_{F}) = -2t~{\rm
cos}(k_y) - \mu$. The fit shows that the form $\Delta E \sim
E_{0}(k_{F})~{\rm cos} \big(2 k_{F} L_{0} + \phi(k_{F}) \big) / L_{0}$,
where $\phi$ is a $k_{F}$ dependent phase and $E_{0}(k_{F})$ is some
energy scale characterizing the electron mediated interaction at the
Fermi energy (but not necessarily of the perturbative RKKY form), works well for all $k_{y}$ modes in our system, even though we are well outside the validity of perturbation theory (as seen in the 1D case
discussed earlier).

%-----------------------------------------------------------------------

The total $\Delta E$ of the quasi-1D system is given by the sum of the
individual terms for each mode $k_{y}$, weighed by the functions
$E_{0}(k_{F})$. Nevertheless, it seems that away from half filling, the
total periodicity is dominated by the value at $k_{y} = 0$, i.e., ${\rm
cos^{-1}}(1 - |\mu| / 2)$. A look at the plots reveals that the curves
flatten out near this point, and their derivatives go to zero. Hence, as
long as $E_{0}(k_{F})$ and $\phi(k_{F})$ are `reasonably' flat functions
of $k_{F}$, the sum will be dominated by terms around $k_{y}=0$, whereas
the rest of the terms will be cancelled quickly due to fast fluctuations
in their phases. In contrast, at $\mu = 0$, the plot is a straight line,
and the sum over $k_{y}$ leads to a flat result without any periodicity
in the regime $L_{0} \ll L_{y}$. This observation immediately explains
the approximate symmetry around $\mu = -2$, as the periods of two
systems with $\mu = -2 -x$ and $\mu = -2 + x$ are the same at $k_{y} =
0$, as confirmed by the plots. On the other hand, since the rest of the
curves are very different, this symmetry is only approximate. 

These observations also provide an intuitive explanation for the contrasting behaviour of the system at small and large $L_{y}$, especially the absence of SP order in the latter regime. Since the overall periodicity at large $L_{y}$ is dominated by a single mode at $k_{y} = 0$, where the system consists of spins $\vec{S}_{i}$ that are oriented identically along the transverse dimension, it is natural to interpret the system in terms of an effective 1D superlattice model, where the long transverse chain of spins is replaced by a single, large, `macro-spin', or equivalently, an enhanced $J_{H}$. However, as discussed in Section \ref{sec:level2}.A, a large $J_{H}$ generically tends to suppress SP phases; thereby, either ferromagnetic or antiferromagnetic orderings are more prevalent in that regime. Of course, the precise details of
how this works in finite systems, with all the competing energy scales at play, are only obtained by numerical means, as our work clearly demonstrates. 

While these arguments provide a basic understanding of several aspects of the results related to the periodicity and ordering in the system, more involved observations, such as the presence of sharp, horizontal `fingers' at $\mu = 0$, cannot be explained by this rudimentary approach.
%% (in the perturbative regime, $E_{0}(k_{F}) =  \Big( \frac{J^{2}_{H}
%% ~k^{2}_{F}}{4 \pi~ {\rm sin}^{2}(k_{F}/2)} \Big)$ (see Appendix
%% \ref{sec:pert}), which increases 

\section{Discussion and Analogy with Experiments}
\label{sec:disc}
%------------------------------------------------------------------------  
\begin{figure}[t]
\centerline{
\includegraphics[scale=0.33]{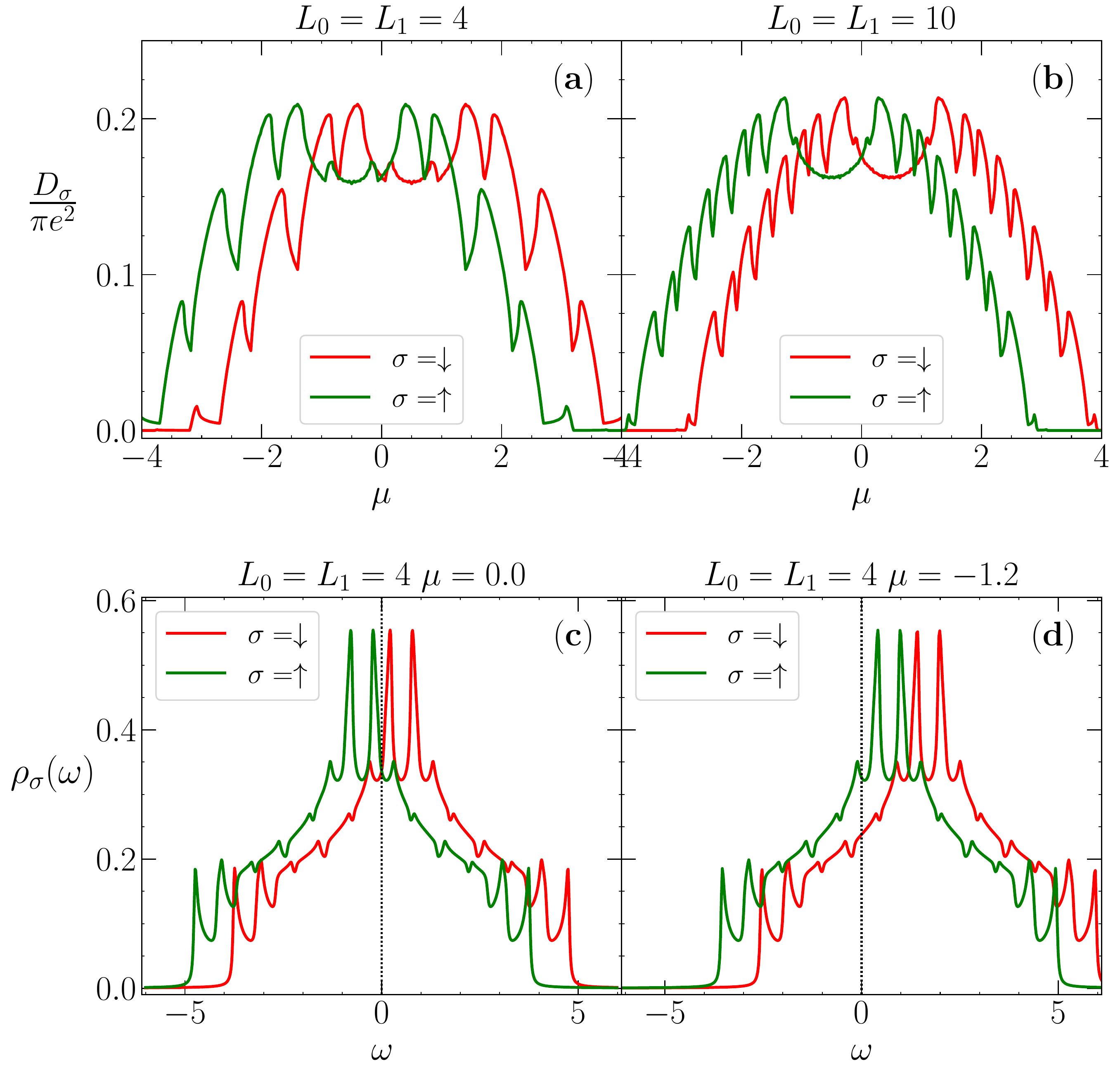} 
}
\caption{Spin resolved Drude weight $D_\sigma$ as a function of $\mu$
for $L_{0} = L_{1} = 4$ (a), and $L_{0} = L_{1} = 10$ (b), for an FM
state, demonstrating finite size oscillations and a concave central
feature of width $\simeq J_{H}$ in both spin channels, leading to higher
value of $D$ for the down channel in this range. Bottom plots show spin
resolved density of states (DOS) $\rho_{\sigma}(\omega)$ at $\mu = 0$
(c) and $\mu = -1.2$ (d), demonstrating (i) only superficial resemblance
with transport due to strong renormalization by current matrix elements
in the transport formula (Appendix~\ref{sec:cond}), (ii) a mirror
symmetry at $\mu = 0$ (left) leading to same transport in both spin
channels, which is no longer present away from half filling (right) (see
text for more details).
}
\label{fig:fig_8}  
\end{figure} 
%-----------------------------------------------------------------------

In this section we discuss several implications of our results and
compare with experimental observations.

\subsection{Mechanism of Transport}

The standard explanation for the phenomenon of GMR states that the
scattering rate of one spin species is different depending on whether
the spin blocks are aligned ferromagnetically or antiferromagnetically.
An effective resistor model based on this can be used to
explicitly show that the FM state has a lower resistivity than the AF
state \cite{Nobel}. 

Since we calculate the transport from first principles using our model,
our results include all the myriad quantum effects stemming from the
finite sizes of $L_{0}$, $L_{1}$, the effect of the transverse modes
$k_{y}$, the spin splitting due to a large, finite $J_{H}$, and so on,
that the heuristic explanation above does not encompass.
Figure~\ref{fig:fig_8} explains this in detail. Figs.~\ref{fig:fig_8}(a) and \ref{fig:fig_8}(b) show the spin resolved
Drude weight $D_\sigma$ (See Appendix~\ref{sec:cond}), as a function of
$\mu$ for fixed values of $L_{y}$ and $J_{H}$, and two different
combinations of $L_{0}$ and $L_{1}$, for a FM ground state (the AF
ground state has the same transport in both spin channels from symmetry
considerations). Figure~\ref{fig:fig_8}(c) plots the spin resolved density
of states (DOS) for comparison. The Drude weights show marked
oscillations with $\mu$, that become finer as $L_{0}$, $L_{1}$ are
increased (top right), indicating finite size effects. The finite
$J_{H}$ results in a pronounced feature near $\mu = 0$ of width $\sim
J_{H}$ in both spin curves. A consequence of this is the highly
non-intuitive result that, for $\mu < 0$, the down spin has a lower resistivity in this regime compared to the up spin. This is a striking example of non-trivial physics due to the combined effect of finite size and a large, finite $J_{H}$, well beyond the scope of perturbative RKKY and scattering approaches. Outside this central region, however, the up spin has the lower resistivity, as expected. A quick look at
Appendix~\ref{sec:cond} reveals that the transport depends on the DOS as
well as the matrix elements $J^{x}_{\alpha \beta}$, and the
the latter could affect the results strongly. A comparison with the DOS
plots in the lower left panel confirms this, where we find a rudimentary
similarity in the oscillations, but the relative strengths are very
different due to the current matrix elements.

The DOS plots [Figs.~\ref{fig:fig_8}(c) and (d)] also highlight another aspect that impacts the transport properties. Due to the particle-hole symmetry in the
system at half-filling, the total density of states is always symmetric
about $E=0$ (See Appendix \ref{sec:ph_symm}). In particular, for the FM
state, where the up and down spins are well defined, this implies that
the spin-resolved density of states, $\rho_{\uparrow}(E)$ and
$\rho_{\downarrow}(E)$, are related by $\rho_{\uparrow}(E) =
\rho_{\downarrow}(-E)$ (as seen in the lower left panel). Hence,  there
is a mirror symmetry in the transitions from filled to empty states in
the current matrix element in Eq.~\eqref{eq:lambda_xx}, and as a result,
the conductivity for both spins is identical. Thus, at half-filling the
increased conductivity in the FM state has a different origin compared
to the standard mechanism. Away from half-filling (bottom right), this
symmetry is no longer present, and the transport can be different for
both spin species, as we have seen before.

This discussion reiterates how our spin-fermion Hamiltonian can capture physical effects in transport related to the geometry, filling, hopping and Hund's coupling that are absent in the standard explanations based on perturbation theory or scattering approaches.
 
 \subsection{Extension to finite $T$}
 
One of the main constraints in our study is the restriction to $T=0$, which ignores the effects of thermal fluctuations in our system.
Before moving to finite temperatures, it is important to discuss whether true long range order can exist in our quasi-1D system at $T>0$, in the presence of effective long range RKKY interactions between the spins $\vec{S}_{i}$. The celebrated Mermin-Wagner (MW) theorem \cite{mw1,mw2,mw3}, which states that it is impossible to have true long range order in isotropic Heisenberg systems at finite temperatures in $d \leq 2$ in the thermodynamic limit, is only valid for sufficiently short-ranged interactions. The situation with long range interactions is far less clear, even though extensions to MW have been conjectured \cite{mw-lr}. If MW retains its validity, then only quasi long range order is possible in such systems. However, for any finite system with a  characteristic length $L$, there exists a finite temperature scale $T_{c}(L)$ below which the spin-spin correlation length exceeds the system size. This sets the nominal critical temperature for that system size. In subsequent discussions of finite temperature order, this interpretation will be implicit.

Ignoring finite temperatures allows us to use variational techniques to explore a large range of parameters and sizes, as we have seen in the previous sections. On the other hand, it implies that even small energy differences can become crucial in determining the ground state and lead to large changes in its properties, an effect that will be washed away readily at finite
temperatures. Our spin fermion model is amenable to sophisticated
techniques for probing finite $T$ effects, including standard Monte
Carlo methods~\cite{dagotto-mc1, dagotto-mc2, dagotto-mc3, weisse-prl, Kipton13,Mondaini14}. However, the requirement of
large dimensions to rule out spiral orders makes it difficult to apply
them in a straightforward manner to our setup due to constraints on
computational time. On the other hand, at low
temperatures, one can perform a simplified statistical analysis using the
spiral mode energies $E(q)$. The expectation value of any fermionic
variable, such as the Drude weight $D$, will then be given by 

\begin{eqnarray}
 \langle D \rangle &=& \frac{\sum_{q} D(q) ~ \exp( -E(q)/T)}{ {\cal Z}}.
\nonumber 
\end{eqnarray}

 An immediate consequence is that at any finite temperature, the
oscillations in $\langle D \rangle$ in Fig.~\ref{fig:fig_4} will decay
with increasing $L_{0}$ due to the diminishing energy difference $\Delta
E$, finally getting completely washed away when $\Delta E(L_{0}) \sim
T$. Similar results will hold for all such observables. 
  
 While it may be difficult to apply Monte Carlo methods to this system
directly, our Hamiltonian may be used to study other heterostructure
devices exploring various magnetic configurations in their constituents,
including tunnel magnetic junctions \cite{kailiu, tmr}, where the dimensions could be small enough to apply the finite temperature techniques discussed above. 
 
 \subsection{Comparison with Experiments}
 
 Our primary result in this paper, illustrated in Fig.~\ref{fig:fig_4},
captures the basic transport properties of GMR by exploiting the
oscillatory nature of the RKKY interactions to demonstrate periodic
variations in the transport with increasing buffer size, as the system
flips between FM and AF states. Several experiments done on a similar
geometry on magnetic-magnetic layers such as ${\rm Co} / {\rm Cr} $ or
${\rm Fe} / {\rm Cr}$ \cite{parkin1}, or magnetic-nonmagnetic layers
such as ${\rm Co} / {\rm Ru}$ and ${\rm Co} / {\rm Cu}$
\cite{parkin1,parkin2}, reported oscillations in the interlayer coupling
and saturation magnetoresistance with changing buffer length. The
increase in magnetoresistance varied between $6 \%$ (${\rm Ru}$ buffer)
to $65 \%$ (${\rm Cu}$ buffer). More recently, experiments on multilayer
van der Waals heterostructures \cite{tmr} have demonstrated giant
Tunneling Magnetoresistance (TMR) of up to $19000 \%$ based on the same
principle of lower resistance in FM configurations. These measurements
were done at finite temperatures, and showed marked attenuation in their
oscillations \cite{parkin1,parkin2}, in contrast to our results. As
discussed before, however, at finite $T$, our model would also yield
similar decaying oscillations in $D$ with increasing buffer size. 

In our current work, we have explicitly neglected the effects of spin-orbit coupling in both our Hamiltonian as well as in the variational ansatz for the $\vec{S}_{i}$. In real materials, such as ${\rm Co} / {\rm Pt}$ or ${\rm NiFe} / {\rm Pt}$ interfaces, the spin-orbit coupling plays an important role due to the absence of inversion symmetry in the system, and results in a canted spin arrangement in the ferromagnetic material \cite{co-pt1, co-pt2, Nembach2015}. In order to model such systems more accurately, it is important to extend the current formalism by including an appropriate Dzyaloshinskii-Moriya term \cite{dm1,dm2}, an interesting direction that we intend to pursue in future work. 

The large variation in magnetoresistance among the different materials indicates a correspondingly large variation in system parameters, even though the underlying principle driving GMR presumably remains the same. Our model provides a broad and thorough framework for
understanding and modeling GMR physics in an array of experimental systems with varying parameters and geometries, and can capture complex physical phenomena beyond the confines of standard perturbation techniques. Possible extensions include modelling one or more specific systems in greater detail by modifying the current Hamiltonian suitably, and studying the ordering and transport in them.

% \textcolor{red}{The experimental results quoted above shows a large variation in the magnetoresistance for different materials and setups, indicating that the system parameters change considerably from system to system, even though the underlying principle driving GMR remains the same. A possible direction for future work would be to concentrate on specific systems in more detail}
 
% Thus, our model provides a broad and thorough framework for
% understanding and modeling GMR \textcolor{red}{physics in an array of experimental systems with varying parameters and geometries, and can capture complex physical phenomena beyond the confines of standard perturbation techniques.} and leaves open the
% possibility of applications to a multitude of more modern
% heterostructures \cite{kailiu, tmr} that exploit the same basic physics.

\section{Conclusions}

In this paper, we explored the ground state and transport properties of a quasi one-dimensional spin-fermion model in a superlattice
geometry. We showed that due to the extended transverse dimension and
superlattice structure, our model may display FM, SP and AF ground states
at different parameter values. However, when $L_{y}$ is much larger than
other dimensions, SP phases are suppressed, and away from half-filling,
the ground state shows periodic oscillations between FM and AF states,
resulting in corresponding oscillations of the transport due to higher
conductivity in the FM state. Using detailed phase diagrams, we
demonstrated how the oscillations varied with $\mu$ and showed that the
basic RKKY form can be used to understand their periodic behavior even
outside the perturbative regime. We discussed the spin resolved transport behaviour with $\mu$, and demonstrated anomalous non-perturbative results that standard scattering theory cannot access. We compared our results to several experiments, establishing the capacity of our model to capture GMR physics comprehensively. The ability of the model to incorporate different geometries and parameter regimes far surpassing conventional perturbative approaches, and access thermal fluctuations using standard Monte Carlo simulations opens up the possibility of studying a range of modern heterostructure devices \cite{kailiu, tmr} where GMR physics is relevant.
% Lastly, by briefly comparing our results to experiments, we outlined
% possible extensions of our work to study different heterostructure
% devices. 

We thank Prof. Elbio Dagotto for useful
comments. ST acknowledges discussions with Stefano Chesi. 
The work of RTS was supported by the grant DE‐SC0014671 funded by
the U.S. Department of Energy, Office of Science.
RM acknowledges
support from NSFC Grants No. U1930402, No. 11674021, 11851110757 and No.
11974039. All numerical computations were carried out on the Tianhe-2JK
at the Beijing Computational Science Research Center (CSRC). 
 
% \bibliography{refs}  
  
\appendix

\section{Derivation of Perturbation Expansion} \label{sec:pert}
In this section, we derive the perturbative expressions used in
Fig.~\ref{fig:1d} (b) and \ref{fig:1d}(c) for the 1D system at
half-filling ($\mu = 0$). When $J_{H} \ll t$, the exchange coupling term
in Eq.~\eqref{eq:hamil} can be treated as a perturbation. Thus, we have
\begin{eqnarray}
 {\cal H} &=& {\cal H}_{0} + {\cal H}_{1}, ~~{\rm where,} \nonumber \\
 {\cal H}_{0} &=& (-t) \sum_{\langle ij \rangle} (c^{\dagger}_{i \sigma} c^{\phantom{\dagger}}_{j \sigma} + {\rm h.c.} ) = (-2t) \sum_{k \sigma} ~{\rm cos}~k~ c^{\dagger}_{k \sigma} c^{\phantom{\dagger}}_{k \sigma} \nonumber \\
 {\cal H}_{1} &=& J_{H} \sum_{i, \alpha, \beta} (\vec{S}_{i}\cdot \vec{\sigma})~ c^{\dagger}_{i \alpha} c^{\phantom{\dagger}}_{i \beta}.
\end{eqnarray}
As in the main text, we assume that the spins 
%% are planar and 
are confined to the $x{\rm -} y$ plane for convenience, so that
$\vec{S}_{i} = ({\rm cos}~ (\theta_{i}),~ {\rm sin}~ (\theta_{i}), ~0)$, where $\theta_{i}$ is the usual polar angle between $\vec{S}_{i}$ and the x-axis. Then
${\cal H}_{1}$ can be written as
\begin{eqnarray}
 {\cal H}_{1} &=& J_{H} \sum_{i, \alpha, \beta} (\vec{S}_{i}\cdot\vec{\sigma})~ c^{\dagger}_{i \alpha} c^{\phantom{\dagger}}_{i \beta} \nonumber \\
 &=& J_{H} \sum_{i} ( e^{- {\rm i} \theta_{i}} c^{\dagger}_{i \downarrow} c^{\phantom{\dagger}}_{i \uparrow} + {\rm h.c.} ) \nonumber \\
 &=& \left( \frac{J_{H}}{N} \right) \sum_{i,k_{1},k_{2}} ( e^{- {\rm i} \theta_{i}} e^{- {\rm i} r_{i} (k_{2} - k_{1})}~ c^{\dagger}_{k_{2} \downarrow} c^{\phantom{\dagger}}_{k_{1} \uparrow} + {\rm h.c.})~~~~~~~~
\end{eqnarray}

To do the perturbation theory at $T=0$, we define the ground state of
the free electron system ${\cal H}_{0}$ by $| G \rangle$, and the
relevant excited states corresponding to an excitation from a filled
state at $ (k_{1}, \sigma)$ with  $k_{1}< k_{F}$ to an empty state
$(k_{2}, -\sigma)$ with  $ k_{2}> k_{F}$ by $| k_{1} \sigma, k_{2}
-\sigma \rangle$.

The first order correction to the energy, $\delta E_{1}$, is given by
\begin{eqnarray}
 \delta E_{1} &=& \langle G | {\cal H}_{1} | G \rangle = 0,
\end{eqnarray}
whereas the second order correction, $\delta E_{2}$, results in
\begin{eqnarray}
 &&\delta E_{2} = - \Big( \frac{J_{H}}{N} \Big)^{2} \sum_{\substack{k_{1} < k_{F} \\ k_{2} > k_{F}}}
 \frac{ \sum_{\sigma}~\big| \langle k_{1} \sigma, k_{2} -\sigma | {\cal H}_{1} | G \rangle \big|^{2}}{-2t ~({\rm cos}~k_{2} - {\rm cos} ~k_{1})} \nonumber \\
 &=&  - \Big( \frac{J_{H}}{N} \Big)^{2} \sum_{\substack{k_{1} < k_{F} \\ k_{2} > k_{F}}}
 \frac{\big\{ \big| \sum_{i} e^{- {\rm i} \theta_{i}} e^{- {\rm i} r_{i}q} \big| ^{2} + \big| \sum_{i} e^{{\rm i} \theta_{i}} e^{- {\rm i} r_{i}q} \big|^{2} \big\}}{-2t~ ({\rm cos}~k_{2} - {\rm cos} ~k_{1})} \nonumber \\
 &=& - \Big( \frac{2 J_{H}}{N} \Big)^{2} \sum_{\substack{ij \\ i<j}} \Bigg\{  \sum_{\substack{k_{1} < k_{F} \\ k_{2} > k_{F}}}
 \Bigg( \frac{{\rm cos} ( r_{ij}q ) ~{\rm cos}\theta_{ij}}{-2t ~({\rm cos}~k_{2} - {\rm cos} ~k_{1})}   \Bigg)
 \Bigg\} \nonumber \\
 &=& - \Big( \frac{2 J_{H}}{N} \Big)^{2} \sum_{\substack{ij \\ i<j}} \Bigg\{  \sum_{\substack{k_{1} < k_{F} \\ k_{2} > k_{F}}}
 \Bigg( \frac{{\rm cos} ( r_{ij} q)  ~\vec{S}_{i}.\vec{S}_{j}}{-2t~ ({\rm cos}~k_{2} - {\rm cos} ~k_{1})}   \Bigg).
 \Bigg\} \nonumber \\
\end{eqnarray}
Here, $r_{ij} = r_{i} - r_{j}$, $q = k_{2} - k_{1}$ and $\theta_{ij} = \theta_{i} - \theta_{j}$. Thus, the pertubative correction is
\begin{eqnarray}
\delta E_{2} &=& - \sum_{\substack{ij \\ i<j}} J_{ij} ~\vec{S}_{i}.\vec{S}_{j}, ~~{\rm where,} \nonumber \\
 J_{ij} &=& 
 \Big( \frac{2 J_{H}}{N} \Big)^{2}   \sum_{\substack{k_{1} < k_{F} \\ k_{2} > k_{F}}}
 \Bigg( \frac{{\rm cos} ( r_{ij} q)}{-2t ~({\rm cos}~k_{2} - {\rm cos} ~k_{1})}   \Bigg). ~~~~~~~~
\end{eqnarray}
 
In the continuum limit, this double sum may be calculated approximately \cite{rkky1, rkky1d} to yield

\begin{eqnarray}
 J(r) &=& \Big( \frac{J^{2}_{H} ~k^{2}_{F}}{4 \pi~ {\rm sin}^{2}(k_{F}/2)} \Big) ~ \Big( \frac{{\rm cos}(2 k_{F} r)}{r} \Big)
\end{eqnarray}

In our superlattice, we may write the total energy as

\begin{eqnarray}
\delta E_{2} &=& - \sum_{\substack{a,m \\ a',m'}} J_{am,a'm'} \vec{S}_{m}.\vec{S}_{m'} \nonumber \\
&=& \sum_{m,m'} \big( \sum_{a,a'} J_{am,a'm'} \big)
\vec{S}_{m}.\vec{S}_{m'}
\end{eqnarray}

Hence, the effective block-block interaction $J_{m,m'} = \big( \sum_{a,a'} J_{am,a'm'} \big)$. By using translation invariance and setting the intra-block couplings $J_{am,a',m}$ to zero (since they only contribute a constant as all $\vec{S}_{i}$ in one block point in the same direction), and using the ansatz for $\vec{S}_{m}$ in Eq. \ref{eq:ansatz}, we have 

\begin{eqnarray}
 &&\delta E_{2} = - \sum_{m,m'} J_{m m'} \vec{S}_{m}.\vec{S}_{m'} \nonumber \\ 
 &=& - \sum_{m m' k} J(k)  ~e^{{\rm i} k (x_{m} - x_{m'})}~{\rm cos} \big(q (x_{m} - x_{m'} ) \big) 
 \nonumber \\ &=& - {\rm Re}~ ( J(q)).
\end{eqnarray}

\section{Conductivity calculation} \label{sec:cond}

In this section, we show the details of the conductivity calculation
used in the main paper, summarized by Eq.(5).

The current operator in the x-direction, $J_{x}$, is given by
\begin{eqnarray}
 J_{x} &=& (-{\rm i} t) \sum_{i,\sigma} \big( c^{\dagger}_{i+x, \sigma} 
c^{\phantom{\dagger}}_{i, \sigma} - {\rm h.c.} \big)
\end{eqnarray}
Using Eqs.~\eqref{eq:ansatz} and \eqref{eq:ft}, this can be written as
\begin{eqnarray}
 J_{x} &=& (-{\rm i} t) \sum_{k_{y}, P, \sigma} \Big\{ \sum^{L_{c} - 1}_{a =
1}  ~ \big( c^{\dagger}_{a+1, k_{y}, P, \sigma} c^{\phantom{\dagger}}_{a, k_{y}, P, \sigma} - {\rm h.c.}  \big)  \nonumber \\ 
 &&+ ~~ ~\big( c^{\dagger}_{1, k_{y}, P, \sigma} c^{\phantom{\dagger}}_{L_{c}, k_{y}, P, \sigma} e^{-{\rm i} P}. - {\rm h.c.} \big) ~\Big\}
\end{eqnarray}

Making use of the general expression for any retarded bosonic operator
\cite{mahan}, the expression for the current current correlation
function $\Lambda_{xx}(q=0, \omega)$ can be written as
\begin{eqnarray}
\label{eq:lambda_xx}
 \Lambda_{xx}(q=0, \omega ) &=& ~ \sum_{nm} \frac{| \langle n | J_{x} | m \rangle |^{2}}{{\cal Z}}~\Bigg( \frac{e^{-\beta E_{n}} - e^{-\beta E_{m}} }{\omega + E_{n} - E_{m} + i \delta} \Bigg), \nonumber \\ \nonumber \\
\end{eqnarray}
where $m$, $n$, are multiparticle eigenstates of the full Hamiltonian, $E_{m}$, $E_{n}$, are the corresponding eigenvalues and ${\cal Z}$ is the partition function.

For a spiral ansatz of the form given by Eq.~\eqref{eq:ansatz}, the
eigenvalues can be calculated by diagonalizing an effective
one-dimensional Hamiltonian given by Eq.~\eqref{eq:hamil_k}. This can be
solved by the following transformation
\begin{eqnarray}
  c_{a,k_{y},P,\sigma} &=& \sum_{\alpha} u^{a}_{\alpha,\sigma} (k_{y},P,q) ~\gamma_{\alpha} (k_{y},P,q).
\end{eqnarray}
Using this transformation, the current operator $J_{x}$ can be written as
\begin{eqnarray}
 J_{x} &=& \sum_{\substack{\alpha \beta \\ k_{y}, P}} J^{x}_{\alpha \beta} ~
\gamma^{\dagger}_{\alpha}(k_{y},P,q) \gamma^{\phantom{\dagger}}_{\beta}(k_{y},P,q)  \mbox{,~where,} \nonumber \\
 J^{x}_{\alpha \beta} &=& (-{\rm i}t) \sum_{\sigma}  \Bigg\{ \sum^{L_{c} - 1}_{a = 1} \Big( u^{\star a+1}_{\alpha \sigma}(k_{y},P,q) u^{a}_{\beta \sigma}(k_{y},P,q) - {\rm h.c.}  \Big) \nonumber \\
 && + \Big( u^{\star 1}_{\alpha \sigma}(k_{y},P,q) u^{L_{c}}_{\beta \sigma}(k_{y},P,q) e^{-i P} - {\rm h.c.} \Big) \Bigg\}.
\end{eqnarray}

Using Eq.~\eqref{eq:lambda_xx} and the fact that the multiparticle states $m,n$ are just Slater determinants of the single-particle eigenstates $| \alpha (k_{y},P,q) \rangle$ (with energies $\epsilon_\alpha$), $\Lambda_{xx}$ can be finally rewritten as
\begin{eqnarray}
 \Lambda_{xx}(q=0,\omega) &=& \sum_{\substack{\alpha \beta \\ k_{y},P}} | J^{x}_{\alpha \beta} |^{2} ~ \Bigg( \frac{f(\epsilon_{\alpha}) - f(\epsilon_{\beta})}{\omega + \epsilon_{\alpha} - \epsilon_{\beta} + i \delta} \Bigg). ~~~~
\end{eqnarray}

The expectation value of the kinetic energy in the $x$-direction,
defined by $k_{x} = \sum_{\langle ij \rangle~ \sigma} \big(
c^{\dagger}_{i \sigma} c_{j \sigma} + {\rm h.c.} \big)$ can be
calculated in exactly the same manner as the current, and using this,
the total and spin resolved (for the FM state) Drude weights are given
by

\begin{eqnarray}
\label{eq:drude_both}
  \frac{D}{\pi e^{2}}&=&\langle -k_{x} \rangle -{\rm Re}~ \Lambda_{xx}(q=0,\omega \rightarrow 0), \ {\rm and} \nonumber \\
   \frac{D_{\sigma}}{\pi e^{2}}&=&\langle -k_{x, \sigma} \rangle -{\rm Re}~ \Lambda_{xx,\sigma}(q=0,\omega \rightarrow 0).
\end{eqnarray}

\section{Particle-Hole symmetry} \label{sec:ph_symm}

In this section, we briefly demonstrate the particle-hole symmetry of
the Hamiltonian at half-filling.  Using the definitions $c = ( c_{1
\uparrow},~.~.~., c_{N \uparrow}, c_{1 \downarrow},~.~.~., c_{N
\downarrow})^{T}$ and ${\bar c} = ( c^{\dagger}_{1 \uparrow},~.~.~.,
c^{\dagger}_{N \uparrow},  c^{\dagger}_{1 \downarrow},~.~.~.,
c^{\dagger}_{N \downarrow})$, the Hamiltonian in Eq.~\eqref{eq:hamil}
can be written compactly as ${\cal H} = {\bar c}~ H ~c$, where  
\begin{equation}
 H =
 \left(\begin{array}{cc} \hat{{\cal H}}_{kin} - J_{H} {\hat S}^{z} & -J_{H} ( {\hat S^{x}} - i {\hat S^{y}} )\\ \\ -J_{H} ( {\hat S^{x}} + i {\hat S^{y}} ) & \hat{{\cal H}}_{kin} + J_{H} {\hat S}^{z} \end{array}\right).
\end{equation}
Here, $(\hat{{\cal H}}_{kin})_{\langle jk \rangle} = (-t)$, where
${\langle jk \rangle}$ denotes a nearest neighbor pair, $({\hat
S}^{z})_{jk} = S^{z}_{j} \delta_{jk}$, etc.

Now, we define a particle hole transformation by the following relations
\begin{eqnarray}
 c &=& U {\bar d},~~{\rm where,} \nonumber \\ \nonumber \\
 U &=& \left(\begin{array}{cc} 
 {\hat 0}   & {\hat M} \\ \\ 
 -{\hat M} & {\hat 0} \end{array}\right), ~~{\rm and,} \nonumber \\ \nonumber \\
 M_{jk} &=& (-1)^{j} \delta_{jk}
\end{eqnarray}

We note that $U$ is unitary (and real), and hence, $U^{\dagger} = U^{T} = U^{-1}$.
 Applying this to the Hamiltonian, we get
 
 \begin{eqnarray}
  {\cal H} &=& {\bar c} H c = d ~U^{-1} H U ~ {\bar d} \equiv d {\tilde H} {\bar d} \nonumber \\
  &=& {\rm Tr}({\tilde H}) - {\bar d} {\tilde H}^{\star} d =  - {\bar d} {\tilde H}^{\star} d
 \end{eqnarray}

 If this transformation is a symmetry, then $ - {\bar d} {\tilde H}^{\star} d = 
  {\bar d} H d$, and, hence, $U^{-1} H^{\star} U = - H$ \cite{ryu}. Now,

  \begin{eqnarray}
   && U^{-1} H^{\star} U \nonumber \\ \nonumber \\
   &=& \left(\begin{array}{cc} 
 {\hat 0}   & -{\hat M} \\ \\ 
 {\hat M} & {\hat 0} \end{array}\right)
 H^{\star}
 \left(\begin{array}{cc} 
 {\hat 0}   & {\hat M} \\ \\ 
 -{\hat M} & {\hat 0} \end{array}\right) \nonumber \\ \nonumber \\
 &=& 
 \left(\begin{array}{cc} M (\hat{{\cal H}}_{kin} + J_{H} {\hat S}^{z}) M & M ( J_{H} ( {\hat S^{x}} - i {\hat S^{y}} ) M \\ \\ M ( J_{H} ( {\hat S^{x}} + i {\hat S^{y}} ) M &  M (\hat{{\cal H}}_{kin} - J_{H} {\hat S}^{z} ) M \end{array}\right)  \nonumber \\ \nonumber \\ 
 &=& 
 \left(\begin{array}{cc} -\hat{{\cal H}}_{kin} + J_{H} {\hat S}^{z} & J_{H} ( {\hat S^{x}} - i {\hat S^{y}} )\\ \\ J_{H} ( {\hat S^{x}} + i {\hat S^{y}} ) & -\hat{{\cal H}}_{kin} - J_{H} {\hat S}^{z} \end{array}\right)
 \nonumber \\ \nonumber \\
 &=& - H .\nonumber \\
  \end{eqnarray}

This proves the particle-hole symmetry of the Hamiltonian at half filling.\\

 %------------------------------------------------------------------------  
\begin{figure}[b]
\centerline{
\includegraphics[scale=0.37]{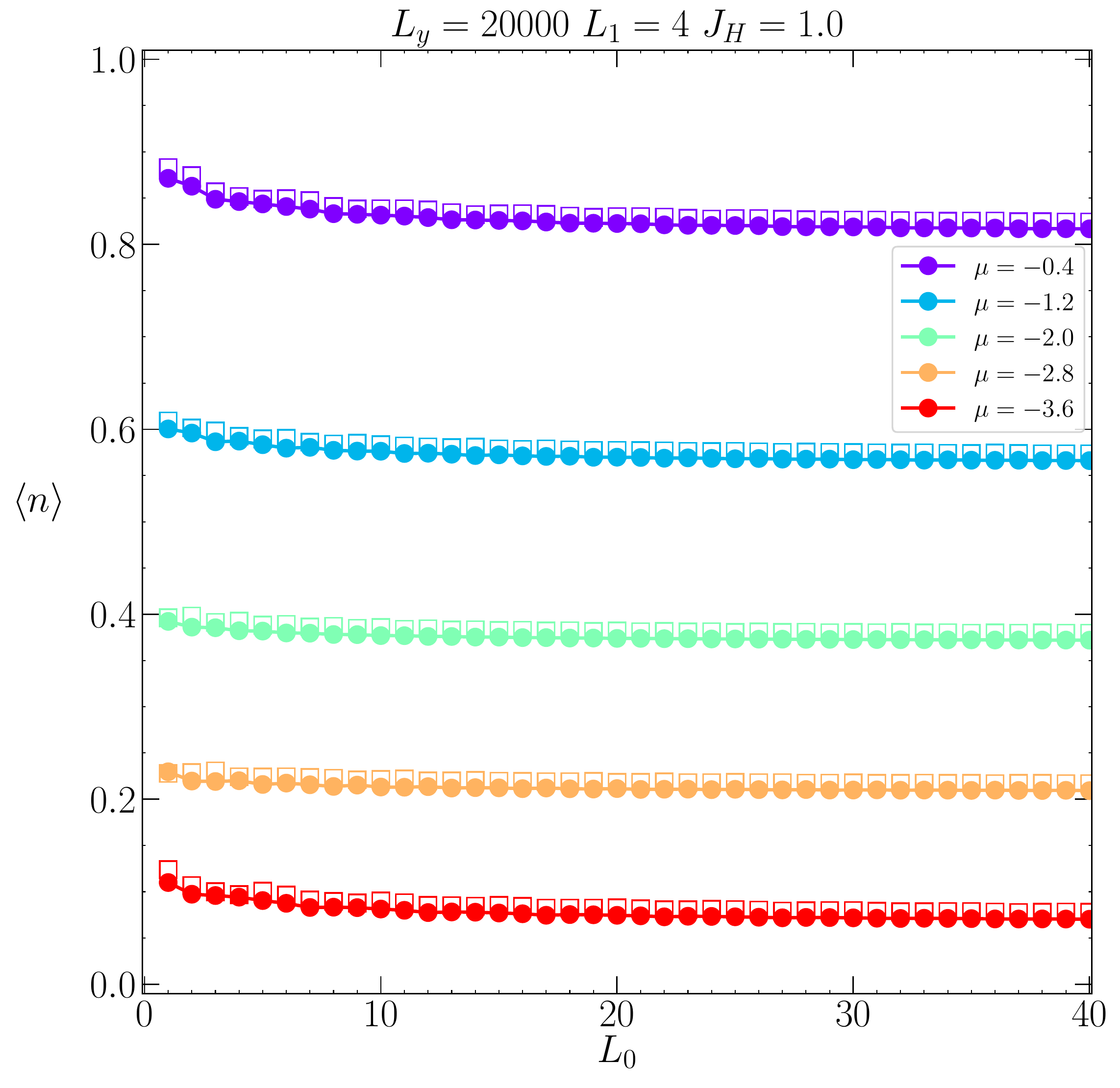} 
}
\caption{The number density $\langle n \rangle$ for the FM state (filled
circles) and the AF state (hollow squares of same color, shifted
slightly for visual clarity) as a function of $L_{0}$ at $\mu = -0.4$,
$-1.2$, $-2.0$, $-2.8$ and $-3.6$ respectively. It is clear that the
variation is marginal ($\lesssim 0.03$) for $L_{0} \gtrsim 5$. As a
result, we choose to do our calculations keeping $\mu$ constant instead
of $\langle n \rangle$.
} 
\label{fig:fig_9}
\end{figure} 
%-----------------------------------------------------------------------

An immediate consequence of this symmetry is that if $| \psi \rangle$
is an eigenstate of $H$ such that $H~| \psi \rangle = E~| \psi \rangle$,
then $U H  ~| \psi \rangle = E U ~| \psi \rangle = U H U^{-1} U~| \psi
\rangle = -H^{\star} U ~| \psi \rangle$. Thus, we have $H U~|
\psi^{\star} \rangle = (-E) U~| \psi^{\star} \rangle$, and hence, for
every eigenstate $| \psi \rangle$ with an eigenvalue $E$, there exists
an eigenstate $U| \psi^{\star} \rangle$ with an eigenvalue $-E$, and the
spectrum is symmetric about zero.\\

%  In our model, we have chosen $J_{H}$ to be positive, but we could
% equally well have chosen it to be negative, since the spectrum is
% independent of the sign of $J_{H}$. It is easy to see this by
% considering a simple unitary transformation $c = U d$, where $U =
% \left(\begin{array}{cc} {\hat 0}   & {\hat I} \\  -{\hat I} & {\hat 0}
% \end{array}\right)$. Then, we have,
%  \begin{eqnarray}
%   {\bar c} H c &=& {\bar d} U^{-1} H U d \nonumber \\ \nonumber \\
%   U^{-1} H U &=&
%   \left(\begin{array}{cc} 
%  {\hat 0}   & {\hat I} \\  
%  -{\hat I} & {\hat 0} \end{array}\right)
%  H
%  \left(\begin{array}{cc} 
%  {\hat 0}   & -{\hat I} \\  
%  {\hat I} & {\hat 0} \end{array}\right) \nonumber \\ \nonumber \\
%  &=& H^{\star}(- J_{H})
%  \end{eqnarray}
% Now, using an argument similar to the one above, we can conclude that
% for every eigenstate $| \psi \rangle$ of $H$ such that $H~|\psi \rangle
% = E~|\psi \rangle$, $U^{-1}|\psi^{\star} \rangle$ is an eigenstate of
% $H(-J_{H})$ with the same eigenvalue.
 
%  \hspace{3cm}
 
\section{Behavior of $\langle n \rangle$} \label{sec:nden}

In this section, we show the behavior of the number density, $\langle n
\rangle$ as a function of the buffer length $L_{0}$ and the chemical
potential $\mu$. Figure~\ref{fig:fig_9} shows the variation of the
average number density $\langle n \rangle$ with the buffer length
$L_{0}$ at different values of the chemica potential $\mu$ (see figure
caption) for the FM and AF states. We see that when $L_{0}$ is not too
small ($L_{0} \gtrsim 5$), the change in density is marginal ($\lesssim
0.03$). Besides, as long as $\langle n \rangle$ is also not too small,
the relative change is rather minor (for example, at $\mu = -2.0$,
$\langle n \rangle$ varies roughly between $0.382$ and $0.372$ for
$L_{0} \gtrsim 5$, a relative change of only about $2.6 \%$. Hence, our
calculations are done at fixed $\mu$, which is much simpler to
implement.
 
%------------------------------------------------------------------------  
\begin{figure}[t]
\centerline{
\includegraphics[scale=0.31]{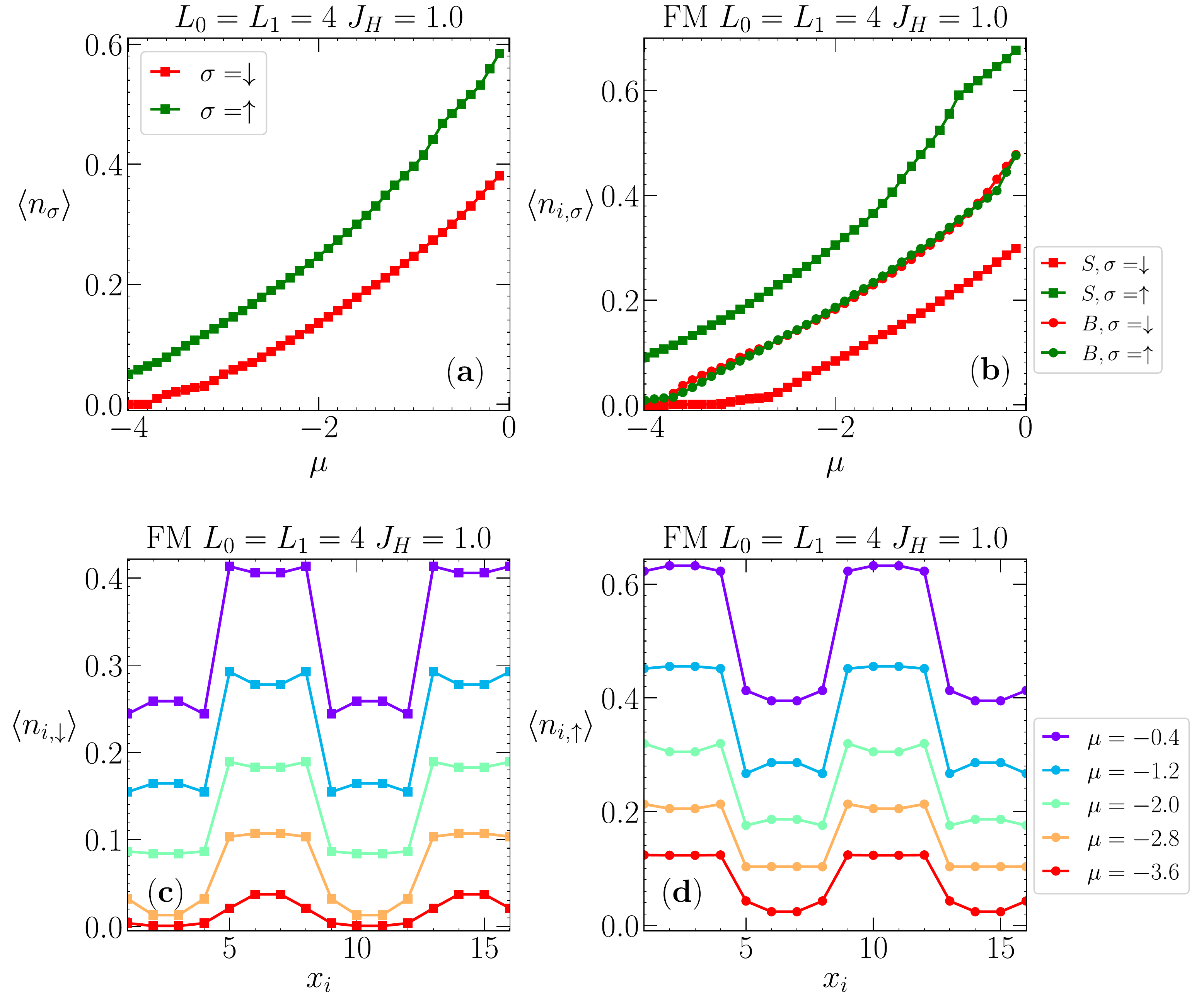} 
}
 \caption{The number density $\langle n_{\sigma} \rangle$ for the FM state as a function of $\mu$ and superlattice co-ordinate $x_{i}$.
(a) plots the total density $\langle n_{\sigma} \rangle$ vs. $\mu$, 
where $\langle n_{i, \uparrow} \rangle > \langle n_{i, \downarrow}
\rangle$ throughout, as expected.
(b) shows $\langle n_{i, \sigma} \rangle$ at the center of the
magnetic ($S$) and buffer ($B$) blocks where we find that the
polarization in the density is largely limited to the $S$ blocks
themselves. (c) and (d) show the variation of $\langle n_{i,\downarrow}
\rangle$ and $\langle n_{i,\uparrow} \rangle$ respectively with the
superblock co-ordinate $x_{i}$ for two superblocks, with expected
oscillations between $S$ and $B$ blocks. Even here, the effect is almost
local.}
\label{fig:fig_10}
\end{figure} 
%------------------------------------------------------------------- 

Figure~\ref{fig:fig_10}, on the other hand, shows the behavior of the
spin resolved densities $\langle n_{i \sigma} \rangle$ with $\mu$ and
the supercell coordinates (consisting of one spin block and one buffer
region respectively) $x_{i}$ for a given set of $L_{0}$, $L_{1}$ and
$J_{H}$ values. Figure~\ref{fig:fig_10}~(a) shows the variation of the
total density $n_{\sigma}$ with $\mu$ for the FM state. As expected,
$\langle n_{\uparrow} \rangle$ is larger throughout the whole range
due to a ferromagnetic $J_{H}$. In comparison, Fig.~\ref{fig:fig_10}~(b)
shows how the local density, $\langle n_{i,\sigma} \rangle$ varies with
$\mu$ at the centre of the magnetic and the buffer blocks (denoted by
$S$ and $B$ in the plots) respectively, for the same FM state. Again, as
expected, $\langle n_{i,\uparrow} \rangle$ is consistently higher in
$S$ (the split is actually bigger here since the total $\langle
n_{\sigma} \rangle$ is the average of the densities in $S$ and $B$). The
densities in the $B$ region, in contrast, are basically identical to
each other, demonstrating that the polarization caused by $J_{H}$ is
largely local.

In Figs.~\ref{fig:fig_10}~(c) and (d), we plot $\langle n_{i,\downarrow}
\rangle$ and $\langle n_{i,\uparrow} \rangle$ as a function of the
supercell coordinates $x_{i}$ for two supercells (for the given
parameters $L_{0} = L_{1} = 4$ that means there are $16$ sites in
total). As expected, $\langle n_{i,\uparrow} \rangle$ ($\langle
n_{i,\downarrow} \rangle$) is larger (smaller) in the $S$ blocks, while
they are essentially the same in the $B$ blocks. As before, we see that
the oscillations between $S$ and $B$ are predominatly local.

\bibliography{ref}

\end{document}